\documentclass[sigconf]{acmart}

\AtBeginDocument{%
  \providecommand\BibTeX{{%
    \normalfont B\kern-0.5em{\scshape i\kern-0.25em b}\kern-0.8em\TeX}}}

\usepackage{amsmath,amssymb,amsfonts}
\usepackage{lipsum}
\usepackage{soul}
\usepackage{adjustbox}
\usepackage{algorithmic}
\usepackage{graphicx}
\usepackage{subcaption}
\usepackage{textcomp}
\usepackage{todonotes}
\usepackage{cleveref}
\usepackage[most]{tcolorbox}
\usepackage[inline]{enumitem}
\usepackage{multirow}
\usepackage{makecell}
\usepackage{algorithmic}
\usepackage{stfloats}
\usepackage{mwe}
\usepackage[font=normalsize]{caption}
\usepackage[font=bf]{caption}
\AtBeginDocument{%
  \providecommand\BibTeX{{%
  Bib\TeX}}}
\usepackage{listings}
\usepackage{url}
\usepackage[normalem]{ulem}
\useunder{\uline}{\ul}{}
\usepackage{xcolor}
\usepackage{hyperref}
\hypersetup{
    colorlinks=true,
    linkcolor={red!50!black},
    citecolor={blue!50!black},
    urlcolor={blue!80!black}
}
\newcounter{observationid}
\newcommand{\defobservation}[2]
{\refstepcounter{observationid}\label{#1}\textbf{Observation~\#\arabic{observationid}:} \textit{#2}}

\setcopyright{cc}
\copyrightyear{2025}
\acmYear{2025}
\setcopyright{cc}
\setcctype{by-nc}
\acmConference[ICPE '25]{Proceedings of the 16th ACM/SPEC International Conference on Performance Engineering}{May 5--9, 2025}{Toronto, ON, Canada}
\acmBooktitle{Proceedings of the 16th ACM/SPEC International Conference on Performance Engineering (ICPE '25), May 5--9, 2025, Toronto, ON, Canada}
\acmDOI{10.1145/3676151.3719372}
\acmISBN{979-8-4007-1073-5/2025/05}

\begin{document}

\title{An Empirical Characterization of Outages and Incidents in \\ Public Services for Large Language Models} 


\author{Xiaoyu Chu}
\email{x.chu@vu.nl}
\affiliation{%
  \institution{Vrije Universiteit Amsterdam}
  \country{The Netherlands}
}

\author{Sacheendra Talluri}
\email{s.talluri@vu.nl}
\affiliation{%
  \institution{Vrije Universiteit Amsterdam}
  \country{The Netherlands}
}

\author{Qingxian Lu}
\email{q.lu@student.vu.nl}
\affiliation{%
  \institution{Vrije Universiteit Amsterdam}
  \country{The Netherlands}
}

\author{Alexandru Iosup}
\email{a.iosup@vu.nl}
\affiliation{%
  \institution{Vrije Universiteit Amsterdam}
  \country{The Netherlands}
}


\renewcommand{\shortauthors}{Xiaoyu Chu, Sacheendra Talluri, Qingxian Lu, \&Alexandru Iosup}
\renewcommand{\shorttitle}{Characterization of LLM Service Outages and Incidents}

\begin{abstract}
People and businesses increasingly rely on public LLM services, such as ChatGPT, DALL·E, and Claude.
Understanding their outages, and particularly measuring their failure-recovery processes, is becoming a stringent problem. 
However, only limited studies exist in this emerging area. 
Addressing this problem, 
in this work we conduct an empirical characterization of outages and failure-recovery in public LLM services.
We collect and prepare datasets for 8 commonly used LLM services across 3 major LLM providers, including market-leads OpenAI and Anthropic.
We conduct a detailed analysis of failure recovery statistical properties, temporal patterns, co-occurrence, and the impact range of outage-causing incidents.
We make over 10 observations, among which: 
(1) Failures in OpenAI's ChatGPT take longer to resolve but occur less frequently than those in Anthropic's Claude;
(2) OpenAI and Anthropic service failures exhibit strong weekly and monthly periodicity; and 
(3) OpenAI services offer better failure-isolation than Anthropic services. 
Our research explains LLM failure characteristics and thus enables optimization in building and using LLM systems.
FAIR data and code are publicly available on \url{https://zenodo.org/records/14018219} and \url{https://github.com/atlarge-research/llm-service-analysis}.
\end{abstract}




\maketitle

\section{Introduction}\label{sec:intro}


\begin{figure}[t]
  \centering
  \includegraphics[width=\linewidth]{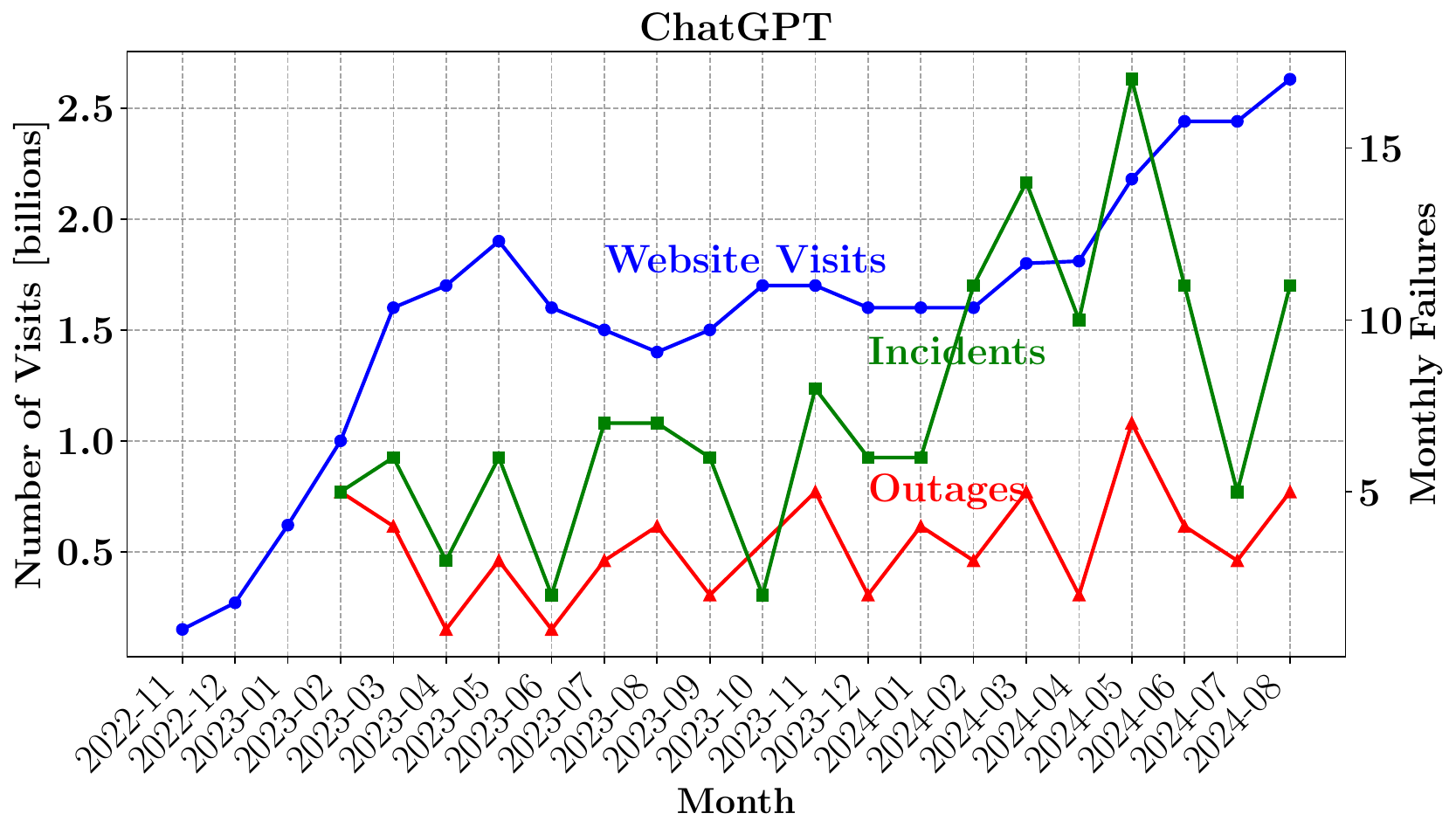}
  \vspace*{-0.7cm}
  \caption{Monthly website visits, outages, and incidents for ChatGPT. Vertical axis: (left) number of website visits in billions; (right) monthly outage and incident counts. Data: website visits \cite{similarwab}, outages (\Cref{tab:outage-dataset}), and incidents (\Cref{tab:incident-dataset}).}
  \label{fig:chatgpt-visits}
    \vspace*{-0.5cm}
\end{figure}

In the past 5 years, increased availability of data and computation enabled Large Language Models~(LLMs) to support scientists, businesses, and general users in a wide range of applications, such as coding \cite{DBLP:conf/chi/Vaithilingam0G22, DBLP:conf/nips/LiuXW023}, image generation \cite{DBLP:conf/nips/KohFS23, DBLP:conf/nips/LuYLWW23}, and general problem-solving \cite{DBLP:journals/tkdd/YangJTHFJZYH24, DBLP:journals/corr/abs-2401-08329}. 
Hundreds of millions of users rely increasingly on public LLM services such as ChatGPT \cite{chatgpt-users-num}, DALL·E \cite{dalle-users-num}, and Claude \cite{claude-users-num}. 
Understanding service outages, and how incidents leading to them are addressed, is essential to enhancing the fault tolerance and quality of service (QoS) of LLM systems.
However, relatively little data and no peer-reviewed studies exist in this rapidly emerging area. 
Addressing this problem, and complementing studies that focus on LLM resource utilization~\cite{DBLP:conf/nsdi/Hu0WWZC0L0L0024, wang2024burstgpt} and user satisfaction~\cite{DBLP:journals/corr/abs-2401-08329},
in this work we conduct the first data-driven, empirical characterization of outages and incidents in public LLM services. We conduct three classes of analysis on long-term datasets we collect from 8 public LLM services from OpenAI, Anthropic, and Character.AI.



The reliability of public LLM services is becoming increasingly important, as service failures can severely erode user experience and cause substantial financial losses under the competitive market.
Driven by demand and market strategy, public LLM providers compete intensely, investing over \$\,40\,billion in 2024~\cite{tr:idc24spending}.
Failures quickly affect many users and become highly visible, as the user cohort has already reached a global scale. For example, the launch of ChatGPT marked a major breakthrough in LLM applications~\cite{fi15060192} and set a user-adoption record, with 100 million monthly active users within 2 months after its 2022 launch~\cite{reuters_chatgpt}. 

Although service reliability is important, users still frequently encounter issues with LLM services. For example, users report to DownDetector many login failures, request errors, and high response latency when using ChatGPT~\cite{downdetector-openai}. 
\Cref{fig:chatgpt-visits} shows the monthly website visits \cite{similarwab}, and outages and incidents for ChatGPT as reported by OpenAI. 
As ChatGPT's monthly web visits grow dramatically, the number of its outages and especially incidents also exhibit an upward trend. 
Thus, significant LLM failures continuously occur, decreasing user satisfaction and potentially causing financial loss, making reliable LLM services a challenge. 

Understanding dependability aspects can help improve systems especially when the workload characteristics are also understood. 
Previous work already provides system-level workload characteristics for the workloads of machine learning~\cite{2023-hotcloudperf-mlfailures, chu2024genericmlworkloadshpc, DBLP:journals/fgcs/VersluisCGLPCUI23}, deep learning~\cite{DBLP:conf/hpca/LiASPABBRBHHHJK22}, big data~\cite{DBLP:conf/wosp/TalluriLAI19}, and more general clouds~\cite{DBLP:conf/cloud/ReissTGKK12}. Recently, LLM workloads have received attention as well~\cite{DBLP:conf/nsdi/Hu0WWZC0L0L0024, wang2024burstgpt}. What remains unaddressed in characterizing the failures of LLM.

We identify and address in this work two main challenges in understanding how public LLM services currently fail. First, \textbf{no longitudinal service failure data currently exists}. Ideally, the community would have access to a large number of similarly curated datasets that capture LLM-service failures, under the same failure model, over long periods of time. 
There are some efforts to provide available LLM workloads, such as BurstGPT~\cite{wang2024burstgpt} and AcmeTrace~\cite{DBLP:conf/nsdi/Hu0WWZC0L0L0024}, but they each focus on one LLM service, and none provides service failure data for it. 
Second, \textbf{no comprehensive analysis of failures in public LLM services currently exists}. 
At this stage in the scientific area, such an analysis would ideally be data-driven, and include for example general characteristics of failures, such as Mean Time Between Failures (MTBF) and To Recovery (MTTR) from classical dependability analysis~\cite{DBLP:journals/tdsc/AvizienisLRL04}, and also of the time spent in various stages of the recovery-process specific to LLM operations;
a temporal analysis of failures;
and
an analysis of failure cascades (co-occurrences). 
Such kinds of analysis would enable future research into models, and future theoretical and practical studies of LLM systems.

Addressing both main challenges, this research aims to provide a thorough empirical characterization of LLM service failures, using data from official outages and incident reports, which are the two types of information self-disclosed by LLM service providers when significant failures occur. 
Our contribution is manyfold:

\begin{enumerate}[label=(\roman*)]
    \item We summarize the de facto industry standard for modeling LLM-service outages and exemplify the anatomy of an outage~(\Cref{sec:tutorial});

    \item We collect outage and incident data for 8\, LLM services, and prepare the corresponding LLM-failure datasets~(\Cref{sec:collection}). This study covers representative, commonly used LLM services, across 3\,LLM-service providers;
    
    \item We analyze the failure characteristics of 8 LLM services~(\Cref{sec:analysis:failure-recovery}). We analyze the MTTR and MTBF by provider and by service, the time spent in various stages of the recovery process, and quantify empirically the model parameters; 
    
    
    \item We analyze LLM service failures over time~(\Cref{sec:analysis:temporal}).
    We explore service availability over hourly and daily intervals, identify various diurnal and weekly patterns, and investigate auto-correlations;
    

    \item We analyze the co-occurrence of failures~(\Cref{sec:analysis:coocurrence}).
    Specifically, we analyze the co-occurrence of failures per provider, and of pairs of services across providers;

    \item We follow the principles of open science and release the datasets and software as open, FAIR artifacts to enable reproducibility and further research.
    
\end{enumerate}

\section{Anatomy of an LLM-service Incident: Model and Example}\label{sec:background}\label{sec:bg}\label{sec:tutorial}

We present in this section a model, coupled with an example, of how an LLM-service incident occurs, affects actual users, and is managed by the LLM-service provider.

\begin{figure}[t]
  \centering
  \includegraphics[width=\linewidth]{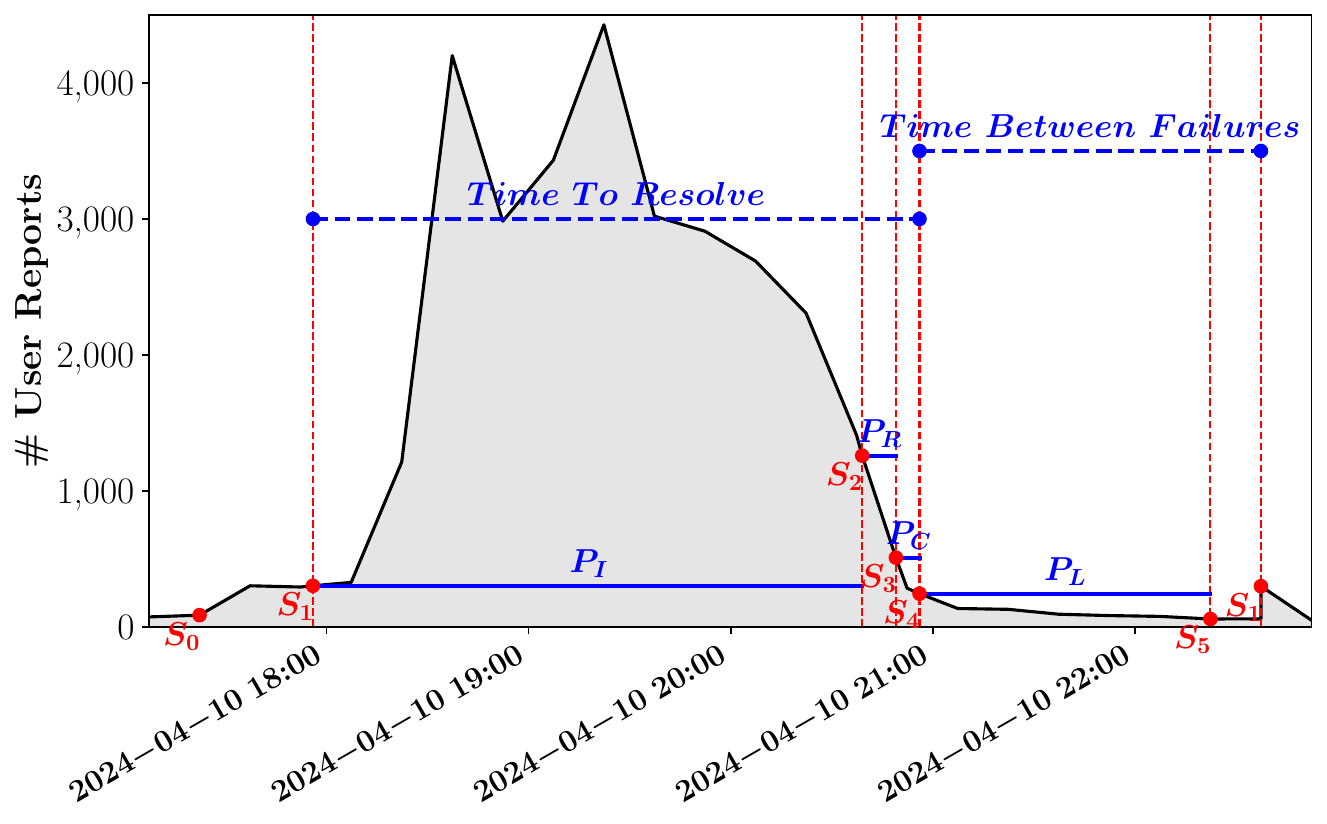}
  \vspace*{-0.7cm}
  \caption{Visualization of the failure-recovery model with user reports of a selected ChatGPT incident, UDT time.}
  \vspace*{-0.5cm}
  \label{fig:visual-fialure-recovery-model}
\end{figure}

\subsection{Model and Real-World Example}\label{sec:model}\label{F-R-modeling}

A failure-recovery process not only leads to addressing a system failure, but also shows a complete story of how the system experienced the (cascading) failure and provides insights to track and improve the system and services affected by it~\cite{DBLP:conf/sc/GamellKKCKP14}.

Industry leads, such as OpenAI and Anthropic, report availability data built around a de facto standard model of their failure-recovery process. Acting as a tutorial, this section summarizes this industry model and exemplifies it through an actual incident. The example considers both data self-reported by the LLM service provider, OpenAI, and data reported by users experiencing the incident; analyzing user-reported data across all incidents studied here is useful but outside the scope of this article.

\textbf{Selection of the incident.}
We selected the real-world incident in which a major outage happened with the ChatGPT LLM service on April 10, 2024; this is the first major incident since, 
on April 1, OpenAI enabled free-to-use access to ChatGPT without signup, effectively opening up ChatGPT trials for everyone \cite{chatgpt-without-sign-up}. OpenAI reported the April 10 incident with complete details about all the stages of its failure-recovery process~\cite{elevated-errors-openai}, which is only done for significant incidents that require many local resources to address. 
In parallel with OpenAI team's efforts to identify and resolve the incident, we recorded two other data sources.
First, users reported problems using the ChatGPT service on DownDetector~\cite{downdetector-archive}; user-reported failures of public services are increasingly used to check the truthfulness and completeness of self-reported failure reports~\cite{DBLP:conf/acsos/TalluriOVTI21}, but so far they have not been used in peer-reviewed studies of LLM services.
We also recorded reports about this outage from news media across the technical and political spectrum, such as 
Fox~\cite{fox5-chatgpt}. 


\begin{table*}[t]
\centering
\caption{Parameters and, below the double line, output metrics of the failure-recovery model proposed in this work.}
\vspace{-0.3cm}
\label{tab:parameters}\label{tab:model}
\resizebox{\textwidth}{!}{
\begin{tabular}{lll}
\toprule
\textbf{ID} & \textbf{Name}  & \textbf{Definition}                                                                                              \\ \midrule
$S_1$           & Investigating Status        & The operational team has started investigating an incident.                                                            \\
$S_2$           & Identified Status          & The issues have been identified.                                                                                  \\
$S_3$           & Monitoring Status          & A fix has been implemented and the operational team started monitoring the results.                                \\
$S_4$           & Resolved Status            & The incident has been resolved.                                                                                   \\
$S_5$           & Postmortem Status          & A summary of the incident after it has been resolved.                                                             \\
\midrule
$P_{I}$ & Investigating Period & From $S_1$ to $S_2$, showing the time from observing to identifying the issues.                          \\
$P_{R}$     & Repairing Period        & From $S_2$ to $S_3$, showing the time to repair the issues.                                          \\
$P_{C}$       & Checking Period      & From $S_3$ to $S_4$, showing the time to confirm the fix is stable and effective.                    \\
$P_{L}$       & Learning Period      & From $S_4$ to $S_5$, showing the time to provide the incident's root cause. \\ 
\midrule
\midrule
$MTTR$ & Mean Time To Resolve & From $S_1$ to $S_4$, covering $P_{I}$, $P_{R}$, $P_{C}$, showing the full time of resolving issues. \\
$MTBF$ & Mean Time Between Failures & From the $S_1$ of the current incident to the next, showing how frequently failures happen. \\
$T$, $T_S$, $A$ & Outage time, scaled, availability & Definitions discussed in \Cref{sec:model:metrics}\\
\bottomrule
\end{tabular}}
\end{table*}

\begin{table*}[t]
\centering
\caption{Values of parameters for the selected incident \cite{elevated-errors-openai}, UDT time. Status-markers $S_1$ through $S_4$ occur on April 10, 2024.}
\vspace{-0.3cm}
\label{tab:selected-incident}
\resizebox{0.89\textwidth}{!}{
\begin{tabular}{ccccccccccc}
\toprule
\textbf{Incident ID} & $S_1$ & $S_2$ & $S_3$ & $S_4$ & $S_5$ & $P_I$ [h] & $P_R$ [h] & $P_C$ [h] & $P_L$ [h] & Time To Resolve [h]\\
\midrule
w20mcckg1748 & 17:56 & 20:39 & 20:49 & 20:56 & 2024-04-18 00:01 & 2.72 & 0.17 & 0.12 & 171.08 & 3.00 \\
\bottomrule
\end{tabular}}
\end{table*}



\textbf{Incident visualization and model parameters:}
Focusing on the major ChatGPT outage on April 10, 2024, \Cref{fig:visual-fialure-recovery-model} visualizes the failure-recovery process as reported by ChatGPT, overlapping it with the number of user reports as reported by DownDetector. Around 2024-04-10 17:30, with the service believed to operate normally, some faults started to happen and the number of user reports increased to an abnormal level. This triggered an alert to the ChatGPT operational team, who started \textit{investigating} at 17:56 (status $S_1$). At 20:39, they \textit{identified} the issue ($S_2$). They quickly implemented a fix, which they released and started \textit{monitoring} at 20:49 ($S_3$). They confirmed that the issue had been \textit{resolved} at 20:56 ($S_4$). During this period, the user reports increased to a peak, around which it fluctuated until the fix was released to increasingly more users, at which point a sharp drop of user reports, toward a normal level, can be seen in the figure. Finally, after a period of \textit{postmortem analysis}, an incident summary was released by the ChatGPT team, to explain the cause of this incident to its users ($S_5$).

This incident exemplifies a failure-recovery process with five key status-markers, $S_1$ through $S_5$. \Cref{tab:model} summarizes the industry-wide model, including these status-markers, the periods they delimit, and the operational metrics. 
Among the periods, whereas $P_{I}$, $P_{R}$, and $P_{C}$ capture how the operational team resolved the outage and are thus synchronous with users experiencing the incident, 
$P_{L}$ captures the period, sometimes ending long after the incident has been resolved, during which the operational team analyses the incident and devises new measures to prevent it and related incidents from happening again.

The model outputs include the industry-standard Time To Resolve and, with knowledge about prior and following incidents, Time Between Failures. \Cref{tab:selected-incident} summarizes the status-markers, the periods spanning the failure-recovery process, and the Time To Resolve for this incident. It shows how the operational team took much longer than usual to find the cause of this incident, 2.72\,h vs. the 0.65\,h found in our analysis in \Cref{sec:analysis:failure-recovery} (see~\Cref{tab:mean-duration}), but then it was able to resolve the failure and restore service much faster than normal. However, by then it was already late, as the media has picked up the incident.

\subsection{LLM-Specific Terms and Metrics}\label{sec:bg:outage-downtime}\label{sec:terms}\label{sec:bg:terms}\label{sec:model:metrics}


\textbf{Incident:}
An operational issue that may cause a service outage, e.g., "getting an error of having reached a limit of GPT-4 usage"~\cite{downdetector-archive}. Once an incident happens, a textual report of its failure-recovery process is produced and is (expectedly) disclosed to the public.
An incident can have different \emph{impact levels} on single or multiple services, which include critical, major, minor, minimal, and maintenance, and are similarly defined by the LLM operators. 

\textbf{Outage:}
Time when the service is unavailable. An outage can have multiple impact ranges: \emph{major outage}, where most of the service's users experience it, and \emph{partial outage}, where a relatively small fraction of the users experience the outage. Operators such as OpenAI \textit{scale} (discount) partial outages as being about 
30\% ($0.3\times$) as bad as major outages~\cite{atlassian-support}.

\textbf{Outage duration:}
For an operator, per day, let $T_M$ be the duration of major outage minutes and $T_P$ be the partial outage minutes.
The formula to calculate the \emph{daily \underline{total} outage minutes ($T$)} is:
\begin{equation}
    T = T_M + T_P
\end{equation}

Similarly, the \emph{daily \underline{scaled} outage minutes ($T_S$)} has the formula: 
\begin{equation}
    T_S = T_M + (T_P \times 0.3)
\end{equation}

Organizations such as OpenAI use primarily the scaled outage minutes to assess and report outage impact~\cite{atlassian-support}.

\textbf{Availability}: 
Derived from the daily scaled outage minutes, 
we define the \textit{daily availability}, $A$, as the percentage of time a service or a group of services are available, given by the formula:
\begin{equation}
    A = (1 - \frac{T_S}{24 \times 60}) \times 100\%
\end{equation}


\begin{table*}[t]
\centering
\caption{Summary of LLM outages, per service. Legend: Incident Count $=$ the number of related incidents.}
\label{tab:outage-dataset}
\vspace{-0.3cm}
\resizebox{\textwidth}{!}{
\begin{tabular}{lllcccrrrrrr}
\toprule
\multirow{2}{*}{\textbf{ID}} & \multirow{2}{*}{\textbf{Service}} & \multirow{2}{*}{\textbf{Provider}} & \multirow{2}{*}{\textbf{Start}} & \multirow{2}{*}{\textbf{End}} & \multirow{2}{*}{\textbf{Months}} & \multicolumn{3}{|c|}{\textbf{Outage Count}}                                                   & \multicolumn{2}{c|}{\textbf{Outage Minutes}} & \multicolumn{1}{c}{\textbf{Incident}} \\ \cmidrule{7-11}
                             &                                   &                                    &                                 &                               &                                  & \multicolumn{1}{|r}{\textbf{Total}} & \textbf{Major} & \multicolumn{1}{r|}{\textbf{Partial}} & \textbf{Total}          & \multicolumn{1}{c|}{\textbf{Scaled}}          & \multicolumn{1}{r}{\textbf{Count}}                                               \\ \midrule
$O_1$                            & API                               & \multirow{4}{*}{OpenAI}            & 2021-02-11                      & 2024-08-31                    & 43                               & \multicolumn{1}{|r}{104}                                & 26             & \multicolumn{1}{r|}{78}                                    & 7,891                   & \multicolumn{1}{r|}{3,340}                    & 242                                                                \\
$O_2$                            & ChatGPT                           &                                    & 2023-02-14                      & 2024-08-31                    & 19                               & \multicolumn{1}{|r}{70}                                 & 28             & \multicolumn{1}{r|}{42}                                    & 5,185                   & \multicolumn{1}{r|}{2,744}                    & 157                                                                \\
$O_3$                            & DALL·E                            &                                    & 2023-02-21                      & 2024-08-31                    & 19                               & \multicolumn{1}{|r}{27}                                 & 13             & \multicolumn{1}{r|}{14}                                    & 2,821                   & \multicolumn{1}{r|}{1,748}                    & 34                                                                 \\
$O_4$                            & Playground                        &                                    & 2021-03-31                      & 2024-08-31                    & 42                               & \multicolumn{1}{|r}{24}                                 & 12             & \multicolumn{1}{r|}{12}                                    & 1,636                   & \multicolumn{1}{r|}{1,018}                      & 36                                                                 \\ \midrule
$A_1$                            & API                               & \multirow{3}{*}{Anthropic}         & 2023-07-11                      & 2024-08-31                    & 14                               & \multicolumn{1}{|r}{25}                                 & 0              & \multicolumn{1}{r|}{25}                                    & 1,675                   & \multicolumn{1}{r|}{502}                      & 80                                                                 \\
$A_2$                            & Claude                            &                                    & 2023-07-11                      & 2024-08-31                    & 14                               & \multicolumn{1}{|r}{30}                                 & 2              & \multicolumn{1}{r|}{28}                                    & 3,017                   & \multicolumn{1}{r|}{983}                      & 90                                                                 \\
$A_3$                            & Console                           &                                    & 2023-07-11                      & 2024-08-31                    & 14                               & \multicolumn{1}{|r}{27}                                 & 1              & \multicolumn{1}{r|}{26}                                    & 2,032                   & \multicolumn{1}{r|}{662}                      & 72                                                                 \\ \midrule
$C_1$                            & Character.AI                      & Character.AI                       & 2023-10-19                      & 2024-08-31                    & 11                               & \multicolumn{1}{|r}{32}                                 & 17             & \multicolumn{1}{r|}{15}                                    & 3,351                   & \multicolumn{1}{r|}{1,878}                    & 41                                                                 \\ \bottomrule
\end{tabular}}
\end{table*}

\begin{table*}[t]
\centering
\caption{Summary of LLM incident reports, per service. Legend: Maint. = Maintenance; Inv. = Investigating; PM = Postmortem.}
\label{tab:incident-dataset}
\vspace{-0.3cm}
\resizebox{\textwidth}{!}{
\begin{tabular}{llccrrrrrrrrrrr}
\toprule
\multirow{2}{*}{\textbf{ID}} & \multirow{2}{*}{\textbf{Provider}} & \multirow{2}{*}{\textbf{First Date}} & \multirow{2}{*}{\textbf{Last Date}} & \multicolumn{1}{|c|}{\textbf{\# of}} & \multicolumn{5}{c|}{\textbf{\# of Impact Levels}}                                                                & \multicolumn{5}{c}{\textbf{\# of Failure-Recovery Status}}                                                   \\ \cmidrule{6-15} 
                             &                                    &                                             &                                            &  \multicolumn{1}{|c|}{\textbf{Reports}}                                         & \textbf{Critical} & \textbf{Major} & \textbf{Minor} & \textbf{None} & \multicolumn{1}{r|}{\textbf{Maint.}} & \textbf{Inv.} & \textbf{Identified} & \textbf{Monitoring} & \textbf{Resolved} & \textbf{PM} \\ \midrule
$P_1$                            & OpenAI                             & 2021-02-09                                  & 2024-08-28                                 & \multicolumn{1}{|c|}{365}                                       & 46                & 125            & 141            & 52            & \multicolumn{1}{r|}{1}                    & 259                    & 144                 & 225                 & 365               & 29                  \\
$P_2$                            & Anthropic                          & 2023-03-25                                  & 2024-08-30                                 & \multicolumn{1}{|c|}{141}                                       & 5                 & 43             & 48             & 44            & \multicolumn{1}{r|}{1}                    & 96                     & 45                  & 51                  & 141               & 2                   \\
$P_3$                            & Character.AI                       & 2023-10-24                                  & 2024-08-07                                 & \multicolumn{1}{|c|}{\ \ 36}                                        & 19                & 11             & 4              & 2             & \multicolumn{1}{r|}{0}                    & 26                     & 16                  & 15                  & 36                & 2                   \\ \bottomrule
\end{tabular}}
\end{table*}

\section{Dataset Collection and  Preparation}\label{sec:collection}\label{sec:data}

We collect for this research long-term datasets from 8 LLM services across 3 LLM service providers. We then process these datasets to prepare data useful to characterize LLM service outages and incidents. 
Tables~\ref{tab:outage-dataset} and~\ref{tab:incident-dataset} summarize the processed outage and incident datasets, respectively.

\subsection{Selection and Introduction of LLM Services}\label{sec:bg:service}

\textbf{Selection process:} 
Addressing the main challenge of lacking longitudinal
failure data about LLM services, particularly under the same failure model, 
%
we carefully investigate the current LLM services, and select 8~LLM services from 3~service providers based on the following reasons: 
(1)~\emph{Data availability:}
Selected services should have public status pages running for long durations, so our data collection can provide rich datasets for the community to further analyze. 
(2)~\emph{Popularity:}
Selected services should be popular, with many users and applications with daily use, so the impact of outages is significant, and there is high likelihood users and media will also report on such outages if left unattended. This pressures operators to respond quickly, so the data we collect represents the best performance LLM operators can currently deliver.
(3)~\emph{Diversity:}
Selected services should cover most types of LLM services provided by different companies. This will ensure the generality of our results.

\textbf{Selected LLM services.}
(1)~\emph{OpenAI API:}
The OpenAI API allows developers to access and use advanced LLM models provided by OpenAI through API keys without building or training from scratch.
(2)~\emph{ChatGPT:}
ChatGPT is a chatbot that interacts with users conversationally. ChatGPT can answer follow-up questions with prompts and provide a detailed response.
(3)~\emph{Labs (DALL·E):}
DALL·E is a text-to-image model that can create original, realistic images from a short text description.
(4)~\emph{Playground:}
Playground is a web-based interface for users to interact with and experiment with OpenAI's language models.
(5)~\emph{Anthropic API:} Similar to OpenAI API, Anthropic API  allows developers to integrate language models such as Claude, into their applications and services.
(6)~\emph{Claude} Similar to OpenAI's ChatGPT, Claude is an AI chatbot and is trained to have natural, text-based conversations with users.
(7)~\emph{Console:} Similar to the OpenAI's playground, the Anthropic Console is a web-based interface that allows users to interact with Anthropic's AI models directly.
(8)~\emph{Character.AI:} is an innovative chatbot platform that leverages LLMs to facilitate a series of chatbots that emulate the personas of various figures, such as historical icons, fictional heroes, modern celebrities, etc.

\subsection{Data Collection and Dataset Preparation}\label{sec:dataset}\label{sec:bg:dataset}

We collected all available outage and incident data reported publicly by of OpenAI, Anthropic, and Character.AI, up to 2024-08-31, on their public status pages~\cite{openai-uptime, anthropic-uptime, characterai-uptime} and incident pages~\cite{openai-incident, anthropic-incident, characterai-incident}.
The starting dates differ: OpenAI has started reporting on February 11, Anthropic on July 11, and Character.AI on October 19, all dates in 2023. Our study misses none of the published reports.

The industry has standardized presenting outage data in a calendar format, with separate information for each service. Each outage history page displays a 3-month calendar view. By hovering over the calendar, one can reveal detailed information about outages, including the occurrence and duration of partial and major outages and any related incidents.
Incident reports provide detailed records of past issues, organized chronologically by month. Each incident report includes a title, a timeline of incident status updates with detailed descriptions, and the services affected.
Not all outages have corresponding incident reports.
Conversely, some incidents, e.g., with minimal impact, do not report a service outage.

We developed an \textit{automated data-collection method}, able to collect industry-standard outage and incident reports. Our tools leverage Python Selenium WebDriver~\cite{selenium}, a robust tool allowing native browser automation by simulating real-user interactions. 
Our tools implement exception-handling mechanisms, addressing potential issues such as network problems, stale elements, and unexpected page layouts. 
They parse and extract information from the dynamic pages and store them as raw outage and incident datasets. 
After that, we performed a series of data transformations for the raw datasets, including filling in missing values, extracting data from text, processing JSON formats, splitting columns, and performing feature calculations to get the metrics used in this study. 

\begin{table}[t]
\centering
\caption{Status counts of incident reports (see~\Cref{tab:incident-dataset}).}
\vspace{-0.3cm}
\label{tab:status-count}
\resizebox{0.3\textwidth}{!}{
\begin{tabular}{cccccrr}
\toprule
\textbf{$S_1$} & \textbf{$S_2$} & \textbf{$S_3$} & \textbf{$S_4$} & \textbf{$S_5$} & \textbf{Count} & \textbf{Percent} \\ \midrule
\checkmark                      &                    & \checkmark                   & \checkmark                 &                    & 131            & 24.39\%             \\
\checkmark                      &                    &                    & \checkmark                 &                    & 110            & 20.48\%             \\
\checkmark                      & \checkmark                  & \checkmark                  & \checkmark                &                    & 77             & 14.34\%             \\
                      &                    &                    & \checkmark                &                    & 62             & 11.55\%             \\
                      & \checkmark                  & \checkmark                  & \checkmark                &                    & 39             & 7.26\%              \\
\checkmark                      & \checkmark                  &                    & \checkmark                &                    & 35             & 6.52\%              \\
                      & \checkmark                  &                    & \checkmark                &                    & 32             & 5.96\%              \\
                      &                    & \checkmark                  & \checkmark                &                    & 18             & 3.35\%              \\
\checkmark                      & \checkmark                  & \checkmark                  & \checkmark                & \checkmark                  & 12             & 2.23\%              \\
\checkmark                      &                    & \checkmark                  & \checkmark                & \checkmark                  & 7              & 1.30\%              \\
\checkmark                      &                    &                    & \checkmark                & \checkmark                  & 5              & 0.93\%              \\
                      &                    &                    & \checkmark                & \checkmark                  & 4              & 0.74\%              \\
                      & \checkmark                  & \checkmark                  & \checkmark                & \checkmark                  & 3              & 0.56\%              \\
                      & \checkmark                  &                    & \checkmark                & \checkmark                  & 2              & 0.37\%              \\ \midrule
\multicolumn{5}{r}{TOTAL}                   & 537            & 100.00\%          \\ 
\bottomrule
\end{tabular}}
\vspace*{-0.3cm}
\end{table}

\begin{figure}[t]
  \centering
  \vspace*{-0.4cm}
  \includegraphics[width=\linewidth]{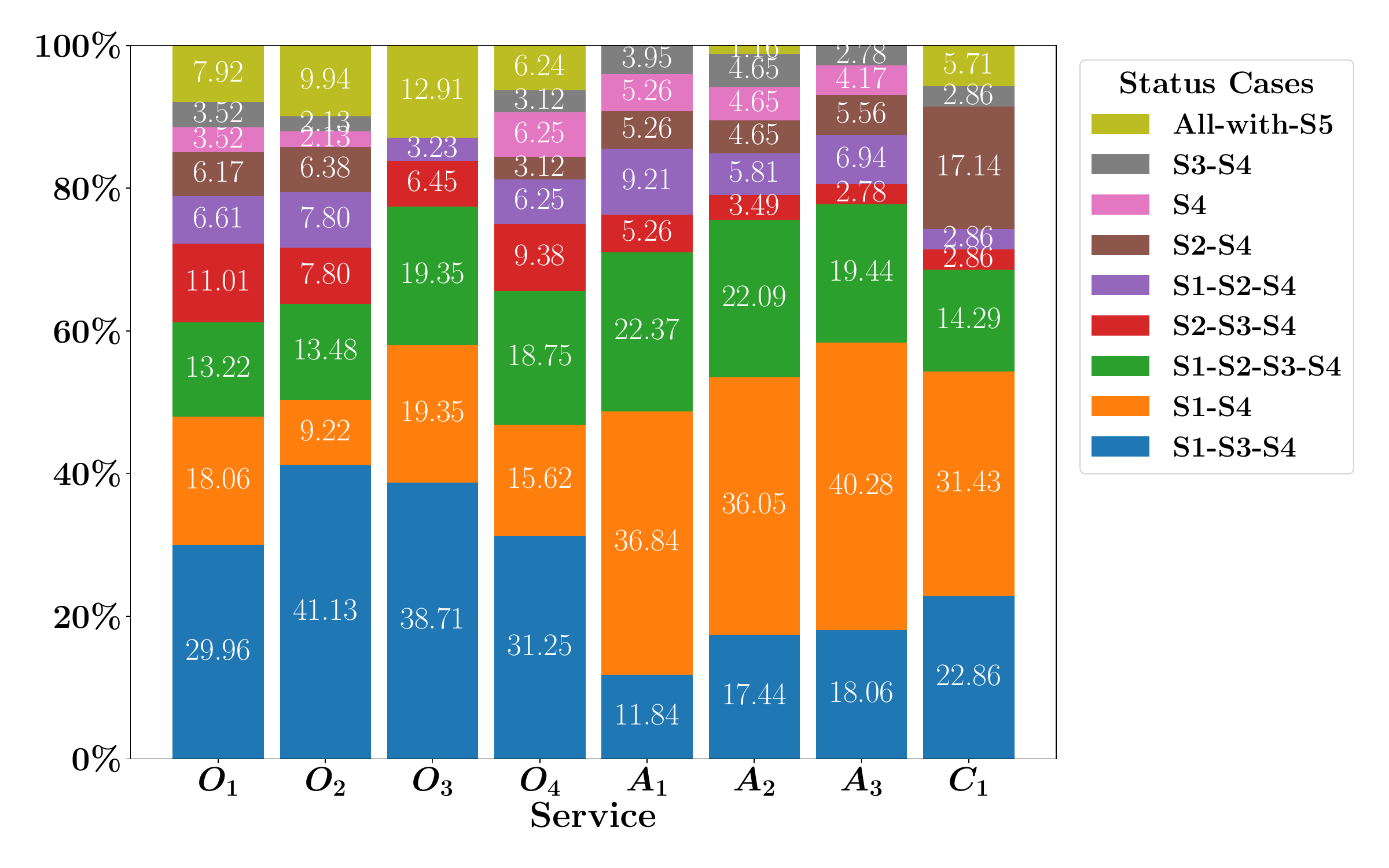}
  \vspace*{-0.8cm}
  \caption{Presence of different status combinations, by service~[\%]. Due to small counts, status combinations with $S_5$ are merged into `All-with-S5'. (\Cref{tab:outage-dataset} indexes the services.)}
  \vspace*{-0.3cm}
  \label{fig:status-combinations}
\end{figure}

\textbf{Data cleanup related to the failure-recovery model in \Cref{sec:model}:}
(1) The incident reports from, e.g., ChatGPT, include 6 statuses: \emph{investigating, identified, monitoring, update, resolved}, and \emph{postmortem.} The \emph{update} status is not considered in the model, but in all the reports we have analyzed it seems to be used only as a keep-alive of the recovery process, to mark the operational team is still actively working on the incident, so we do not consider it in our analysis; 
(2) Out of the over 500 hundreds of incidents we analyzed in this work, only 5 cases do not follow the order of status markers $S_1$ through $S_5$. In 2 of these cases, the status-marker \emph{identified} comes before \emph{investigating}, and in 3 other cases, the status-marker \emph{monitoring} comes before \emph{identified}. None of these cases involves unusual durations or recovery times, so we safely exclude these 5 corner cases in our analysis.



\section{Failure-Recovery Analysis}\label{sec:failure-recovery}\label{sec:analysis:failure-recovery}

We investigate the time spent on the key operational metrics (MTTR, MTBF) and compare the failure-recovery performance across the 8 LLM services and 3 service providers.
We conduct several types of analysis to investigate the failure-recovery processes of LLM services: (1) Statues count and percent of different services; (2) Mean values for main model parameters; (3) Percent of different periods in the MTTR process; (4) Distribution of MTTR and MTTF duration by service; (5) Distribution of MTTR and MTTF duration by providers.
For each analysis, we begin with key observations, followed by detailed descriptions and discussions.

\noindent\fbox{%
\parbox{\linewidth}{%
\defobservation{ob:status-count}{Most incident reports lack information for all statuses. Although 100\% of the incidents have been resolved, only 6.15\% of reports disclose a postmortem.}
}}


\begin{table}[t]
\caption{Mean value for model parameters by service. Legend: h$=$hour(s), D$=$day(s).}
\vspace{-0.3cm}
\label{tab:mean-duration}
\resizebox{\linewidth}{!}{
\begin{tabular}{llrrrrrr}
\toprule
\textbf{ID} & \textbf{Service}     & \textbf{$P_I$ [h]} & \textbf{$P_R$ [h]} & \textbf{$P_C$ [h]} & \textbf{$P_L$ [D]} & \textbf{$MTTR$ [h]} & \textbf{$MTBF$ [D]} \\ \midrule
$O_1$        & API-OpenAI           & 0.72          & 1.63          & 1.46          & 4.10           & 2.56          & 5.64          \\
$O_2$        & ChatGPT              & 0.65          & 1.64          & 1.73          & 4.79          & 3.64          & 4.01          \\
$O_3$        & DALL·E               & 1.01          & 0.96          & 1.81          & 1.86          & 3.03          & 18.24         \\
$O_4$        & Playground           & 0.37          & 1.56          & 2.22          & 4.30           & 2.95          & 39.93         \\ \midrule
$A_1$        & API-Anthropic        & 1.04          & 1.11          & 1.37          & -             & 2.81          & 5.22          \\
$A_2$        & Claude               & 1.35        & 1.72          & 2.05          & 0.21          & 3.16          & 4.79          \\
$A_3$        & Console              & 0.94          & 0.34          & 0.58          & -             & 2.05          & 5.73          \\ \midrule
$C_1$        & Character.AI         & 0.40           & 0.50           & 1.73          & 3.61          & 3.95          & 8.74          \\ \midrule
            & \textbf{Arith. Mean} & 0.84          & 1.40           & 1.58          & 4.01          & 2.94          & 7.41          \\
            & \textbf{Geom. Mean}  & 0.53          & 1.15          & 0.87          & 3.45          & 3.99          & 3.26          \\ \bottomrule
\end{tabular}}
\vspace{-0.6cm}
\end{table}

\begin{figure}[t]
  \centering
  \includegraphics[width=\linewidth]{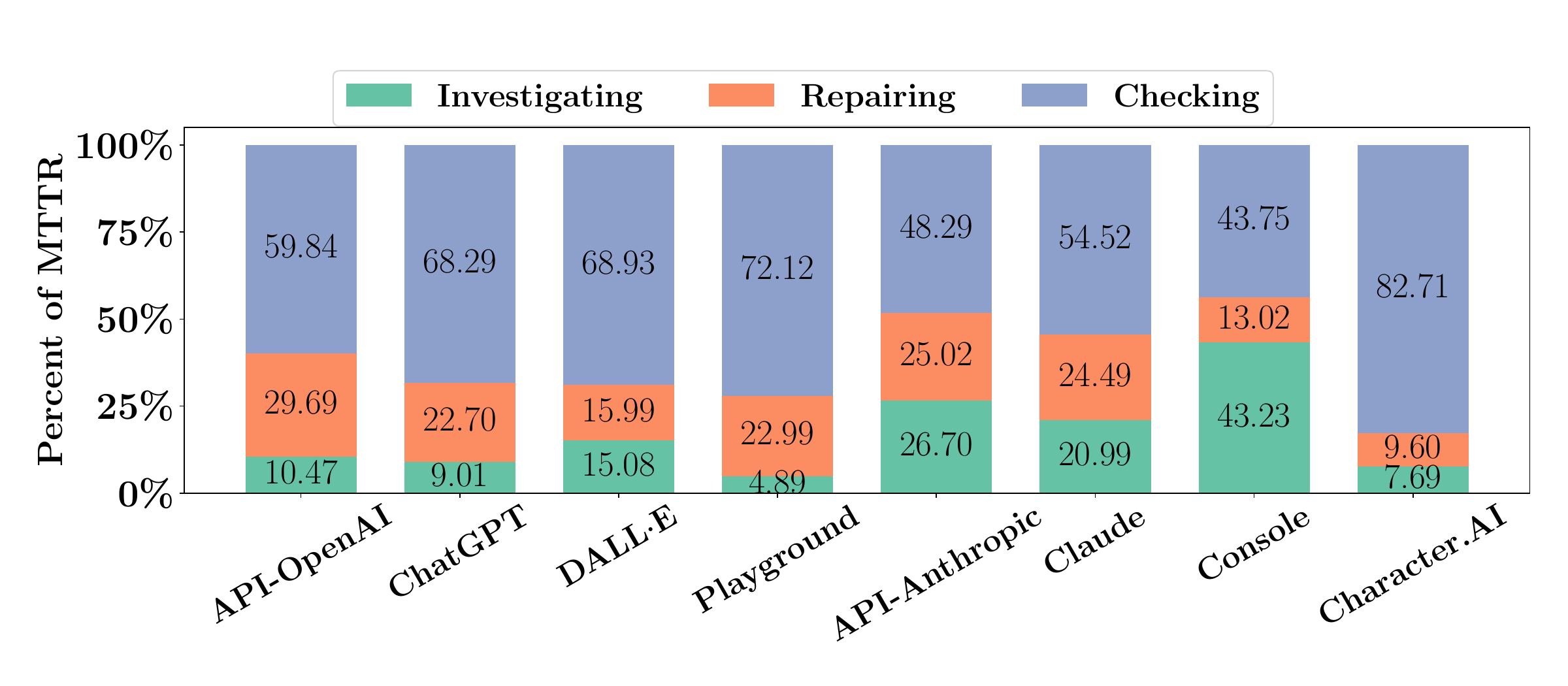}
  \vspace*{-0.9cm}
  \caption{Percent of time spent in the Investigating, Repairing, and Checking periods, from the overall duration for failure resolution [\%].}
  \vspace*{-0.5cm}
  \label{fig:percent-MTTR}
\end{figure}

Updated status information is important for users waiting for a service to recover so that they can plan their work, recovery, and communication with customers.
\Cref{tab:status-count} gives the numbers and percent of different combinations of statuses for all reports. In most cases, incident reports do not include every status.
Cases with $S_5 (postmortem)$ account for the fewest percent, with only 6.15\% of incident reports having a postmortem.
The most prevailing case combination is $S_1$-$S_3$-$S_4$ (24.39\%), in which the important status $S_2 (identified)$ is missing. This means the operators do not communicate to the users that they have identified the issue.
20.48\% of cases only provide information about $S_1 (investigating)$ and $S_4 (resolved)$ statuses.
14.34\% of reports update every status ($S_1$-$S_2$-$S_3$-$S_4$) throughout the duration of incidents, while 11.55\% only update once the incidents have been resolved ($S_4$).

These results indicate that operators fail to communicate the state of a failure to the user and update the user about potential fix times. Users must create their own failure models~\cite{DBLP:conf/dsn/GargPCT18} and fault-tolerance systems~\cite{DBLP:conf/nsdi/PrimoracAB21} and expect little input from the operator. 

\begin{figure*}[t]
  \centering
  \begin{minipage}{\textwidth}
  \begin{subfigure}[b]{0.5\textwidth}
    \includegraphics[width=\linewidth]{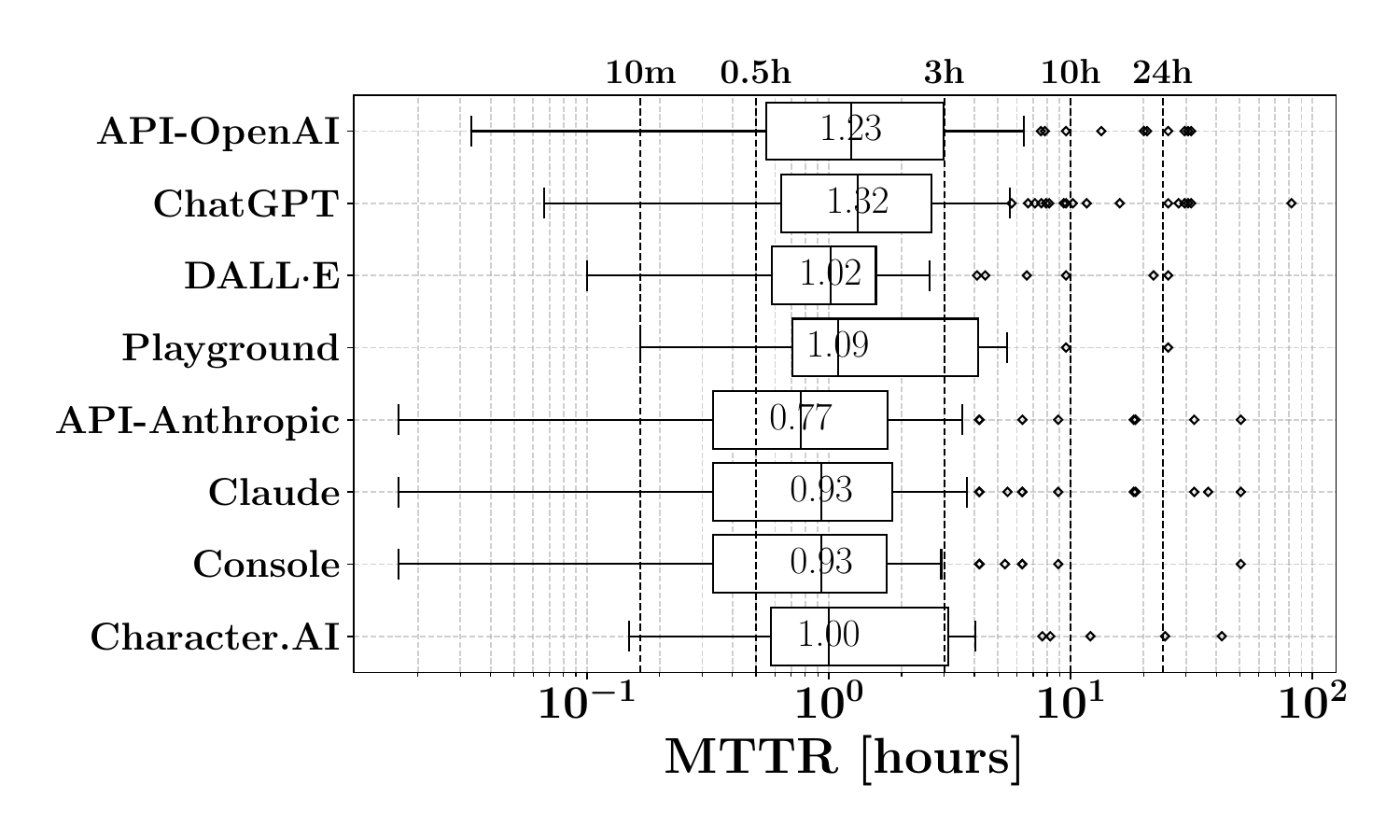}
        \vspace*{-0.7cm}
    \caption{Mean Time To Resolve (Shorter is better).}
    \label{fig:mttr-boxplot}
  \end{subfigure}
  \begin{subfigure}[b]{0.5\textwidth}
    \includegraphics[width=\linewidth]{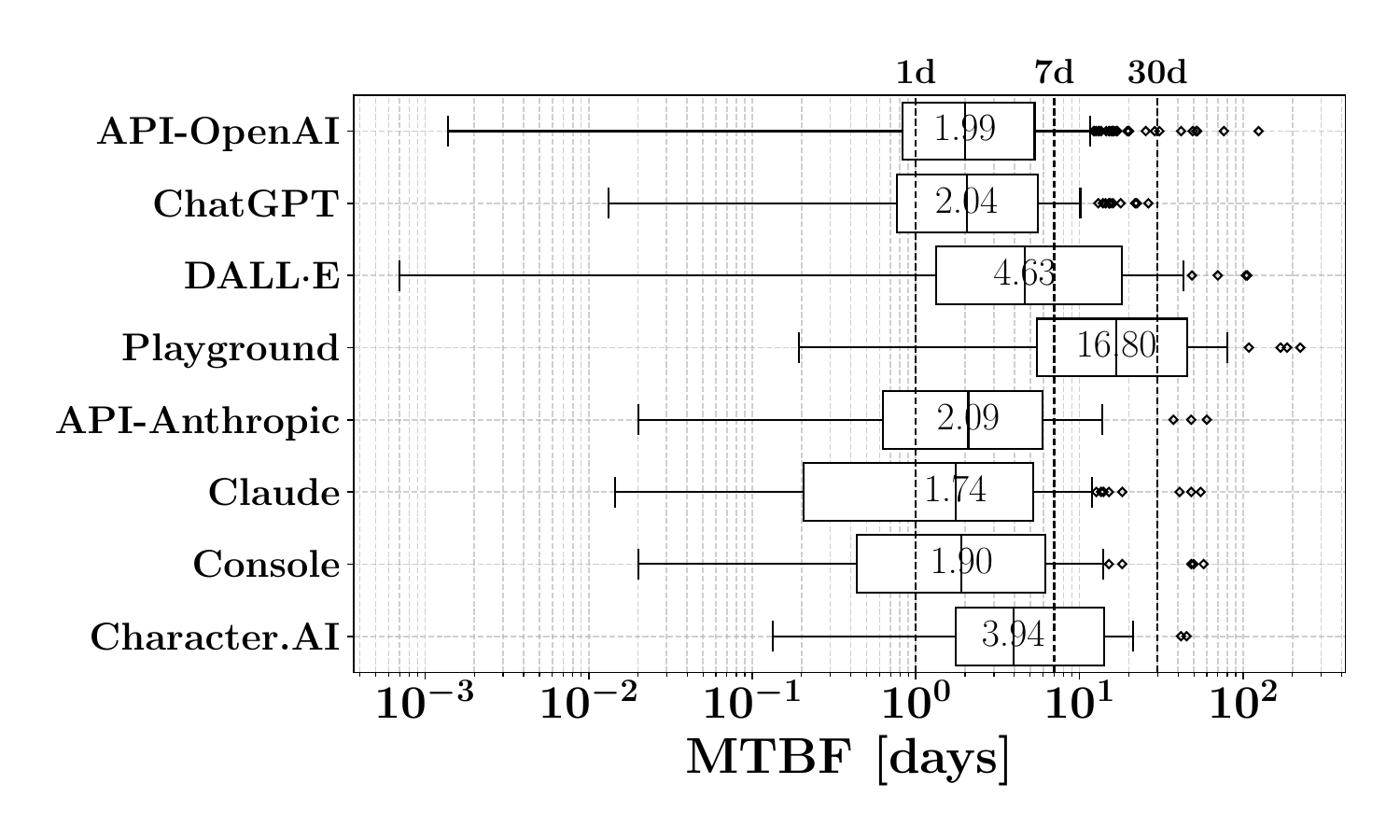}
        \vspace*{-0.7cm}
    \caption{Mean Time Between Failures (Longer is better).}
    \label{fig:mtbf-boxplot}
  \end{subfigure}
  \end{minipage}
  \vspace{-0.3cm}
  \caption{Distribution of MTTR and MTBF by service, with median values indicated.}
  \vspace{-0.3cm}
  \label{fig:mttr-and-mtbf-boxplot}
\end{figure*}

\noindent\fbox{%
\parbox{\linewidth}{%
\defobservation{}{
ChatGPT includes postmortems in 12.91\% of its reports.
Anthropic's reports contain the fewest postmortems, with none provided for its API and Console services.}
}}

The status combinations also vary depending on different services, as \Cref{fig:status-combinations} shows.
The primary status combination for OpenAI services is $S_1$-$S_3$-$S_4$, representing 29.96\% for API, 41.13\% for ChatGPT, 38.71\% for DALL·E, and 31.25\% for Playground.
In contrast, Anthropic and Character.AI primarily use the $S_1$-$S_4$ combination, accounting for over 30\% of each service.
OpenAI publishes more status information in incident reports compared to Anthropic and Character.AI.

\vspace*{0.06cm}
\noindent\fbox{%
\parbox{\linewidth}{%
\defobservation{ob:mean-period-duration}{Claude spent the longest time on investigating (1.35 hours), repairing (1.72 hours), and checking (2.05 hours).}
}}


The time a service takes to resolve incidents affects the fault-tolerance strategies a user can use. For example, users can maintain a local cache to tolerate very short failures. \Cref{tab:mean-duration} shows the mean value of different model parameters. For MTTR and MTBF, because some records don't have $investigating$ timestamps, we use the minimum ones before the resolved status here (This also explains why Claude has the longest $P_I$, $P_R$, and $P_C$ respectively, but it does not have the longest $MTTR$).
The learning period ($P_L$) takes the longest time (2.94 days) in all services. 

\vspace*{0.06cm}
\noindent\fbox{%
\parbox{\linewidth}{%
\defobservation{ob:percent-MTTR}{Significant differences are observed across the 8 services in the percentage of periods within failure resolutions. For Character.AI, 82.71\% of the time is spent on monitoring and checking if the fix is stable and effective. Anthropic services spent more percent of the time for investigating and resolving issues than OpenAI services.}
}}


\Cref{fig:percent-MTTR} shows the percent of the 3 periods (Investigating, Repairing, Checking) in $MTTR$. 
The majority of the resolution time is used for checking, ranging from the highest 82.71\% for Character.AI to the lowest 43.75\% for Console.
Most services spent more percent of the time on repairing than investigating, except for API and Console from Anthropic. Anthropic API spent 26.70\% of the time investigating issues, while Console spent 43.23\% in the same period.
The large fraction of the time used for checking indicates that deploying a fix to production takes a long time. Operators should employ faster testing and continuous deployment techniques to deploy fixes faster~\cite{DBLP:conf/sigsoft/ZhangVWF18}. However, this is challenging as LLMs are a new technology, and there isn't much work on improving testing and deployment time.

\noindent\fbox{%
\parbox{\linewidth}{%
\defobservation{obs:MTTR}{
OpenAI's API and ChatGPT recover slower from failures than Anthropic's API and Claude, with 1.6x and 1.4 longer $MTTR$, respectively.}
}}
\vspace*{0.03cm}

\Cref{fig:mttr-boxplot} depicts $MTTR$ of the 8 LLM service incidents, showing how quickly failures are resolved.
Most failures are resolved between 0.5 and 3 hours, with the median values around 1 hour.
APIs and chatbots are the most popular LLM services. The median $MTTR$ of OpenAI API (1.23 hours) is 1.6x longer than Anthropic API (0.77 hours). Similarly, the median $MTTR$ of ChatGPT (1.32 hours) is 1.4x longer than Claude (0.93 hours).

\begin{figure}[t]
  \centering
  \vspace*{-0.3cm}
  \begin{minipage}{\textwidth}
  \begin{subfigure}[b]{0.25\textwidth}
    \includegraphics[width=\linewidth]{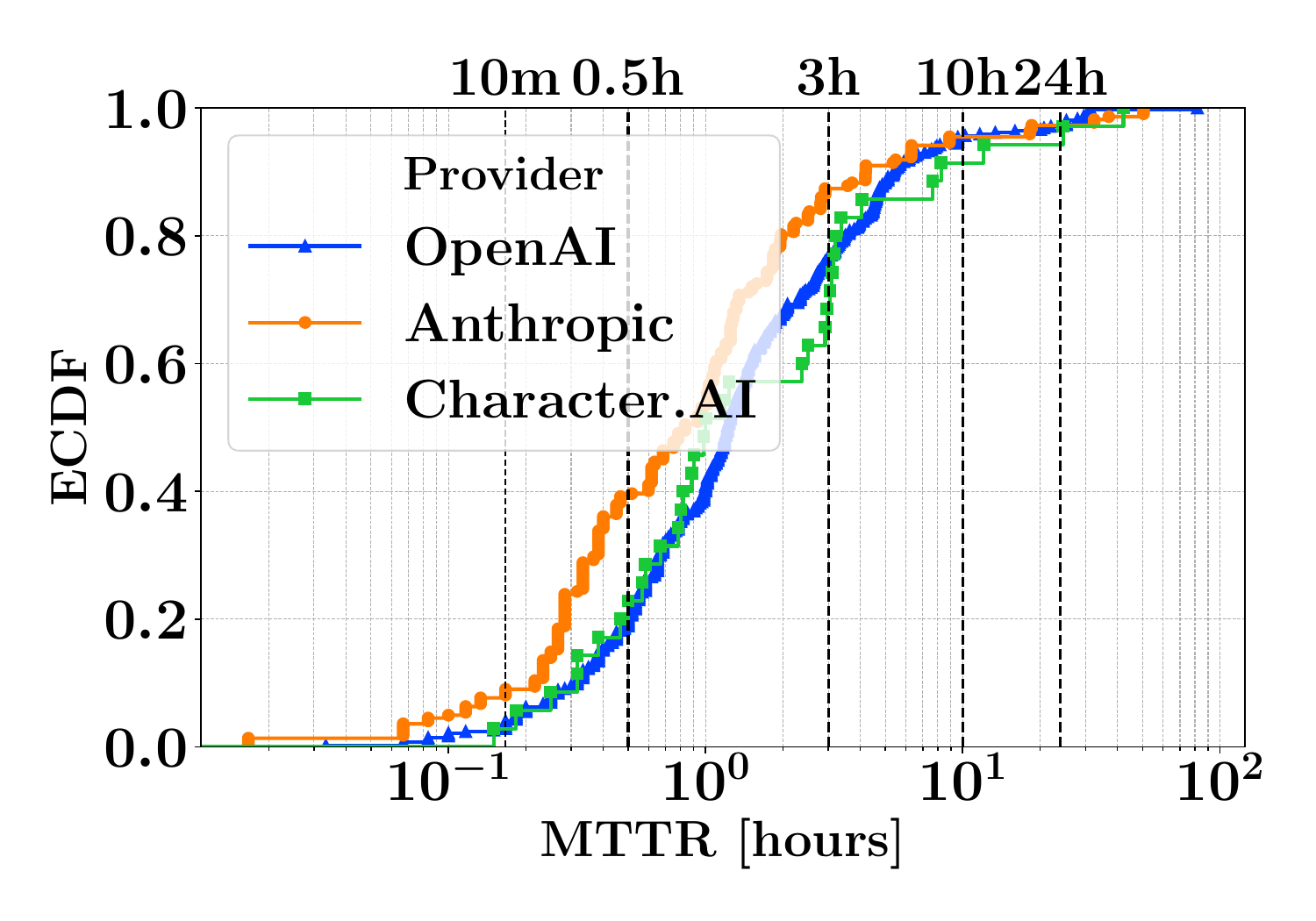}
        \vspace*{-0.7cm}
    \caption{MTTR.}
    \label{fig:mttr-provider}
  \end{subfigure}
  \hspace{-0.4cm}
  \begin{subfigure}[b]{0.25\textwidth}
    \includegraphics[width=\linewidth]{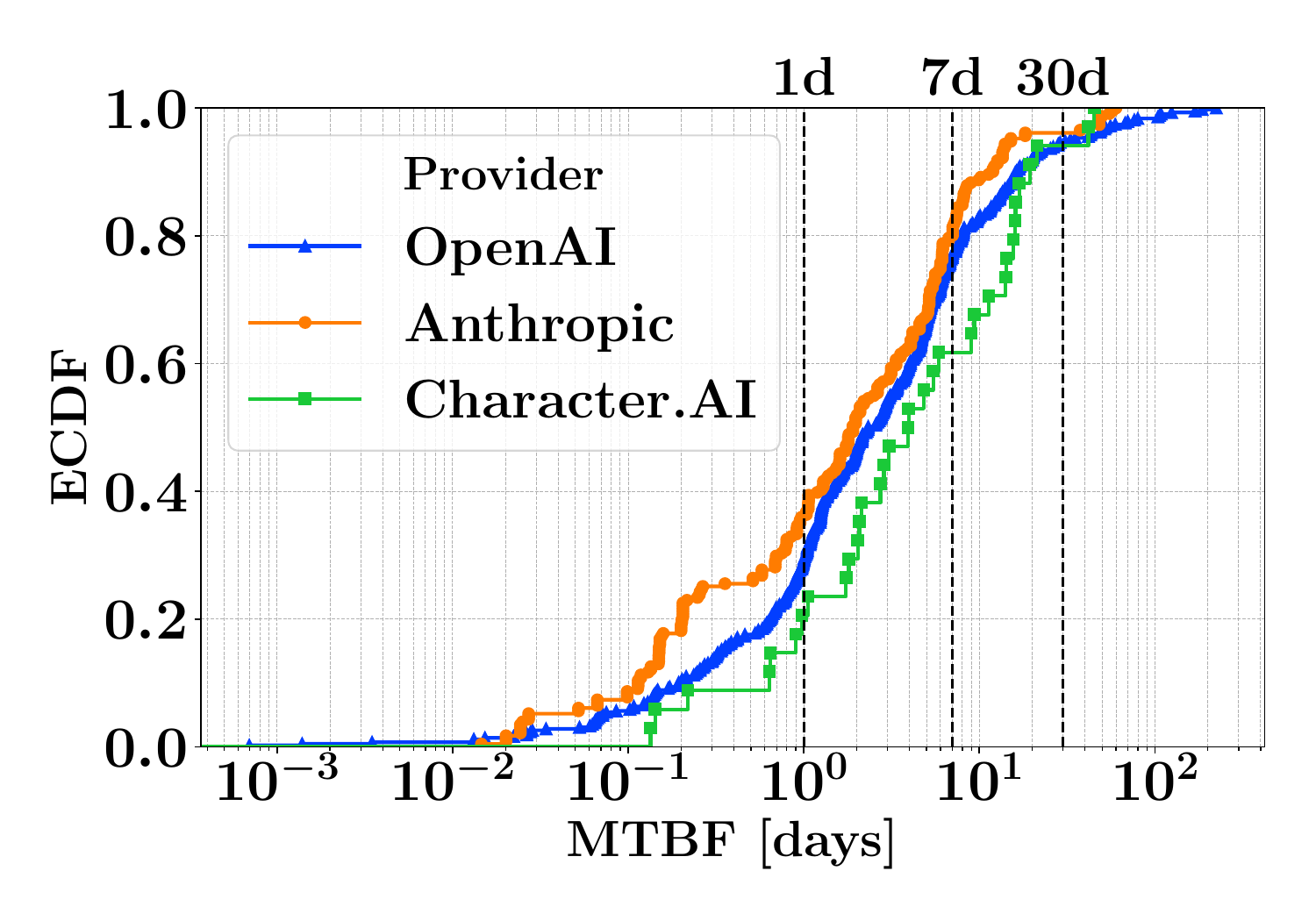}
        \vspace*{-0.7cm}
    \caption{MTBF.}
    \label{fig:mtbf-provider}
  \end{subfigure}
  \end{minipage}
  \vspace{-0.3cm}
  \caption{MTTR and MTBF by provider, ECDF plot. The closer the line is to the upper left, the shorter the time it takes.}
  \vspace*{-0.8cm}
  \label{fig:mttr-and-mtbf-provider}
\end{figure}

\vspace*{0.06cm}
\noindent\fbox{%
\parbox{\linewidth}{%
\defobservation{obs:MTBF}{Playground is the most reliable service (16.80 median $MTBF$), followed by DALL·E (4.53 median $MTBF$) and Character.AI (3.94 median $MTBF$). OpenAI's ChatGPT is more reliable than Anthropic's Claude, though its API is less reliable in comparison.}
}}
\vspace*{0.03cm}

Awareness of how frequently a service fails is important for users to assess the reliability they can offer when they depend on the service. It's also important to assess which fault-tolerance mechanisms they should use as each has a different overhead. For example, active replication, frequent checkpointing, or infrequent checkpointing. \Cref{fig:mtbf-boxplot} depicts $MTBF$ of the 8 LLM incidents, showing how frequently failures occur.
The $MTBF$ of failures varies significantly across services.
The most reliable service is Playground, with a median $MTBF$ of 16.80 days, which is nearly 9.66 times higher than the lowest median $MTBF$ of 1.74 days from Claude.
The median $MTBF$ for OpenAI's API is 1.99 days, which is lower than Anthropic's API at 2.09 days; however, ChatGPT at 2.04 days is higher than Claude's 1.74 days.
DALL·E and Character.AI are relatively reliable services, with median $MTBF$ values of 4.63 days and 3.94 days, respectively.

The MTBF of LLM services (4-40 days) is much higher than the MTBF of single-node failure (0.25-1 day) and system-wide failure (6.6 days) in other large-scale systems~\cite{DBLP:conf/dsn/MartinoKIBFK14, DBLP:conf/sc/GuptaPET17}. This indicates that LLM operators use effective fault-tolerance mechanisms. It also indicates that users can use low-overhead fault-tolerance techniques like infrequent checkpointing to provide a reliable service that depends on LLMs.

\vspace*{0.06cm}
\noindent\fbox{%
\parbox{\linewidth}{%
\defobservation{obs:mttr-provider}{Over 90\% of the incidents end within 10 hours for all measured providers. Specifically, Anthropic's services resolved failures more quickly but also experienced the highest frequency of incidents, based on its $MTTR$ (2.70 hours) and $MTBF$ (5.22 days) on average.}
}}
\vspace*{0.03cm}

\begin{figure*}[t]
    \centering
    \begin{minipage}{\linewidth}
    \begin{subfigure}[b]{\linewidth}   
        \centering 
        \includegraphics[width=0.82\linewidth]{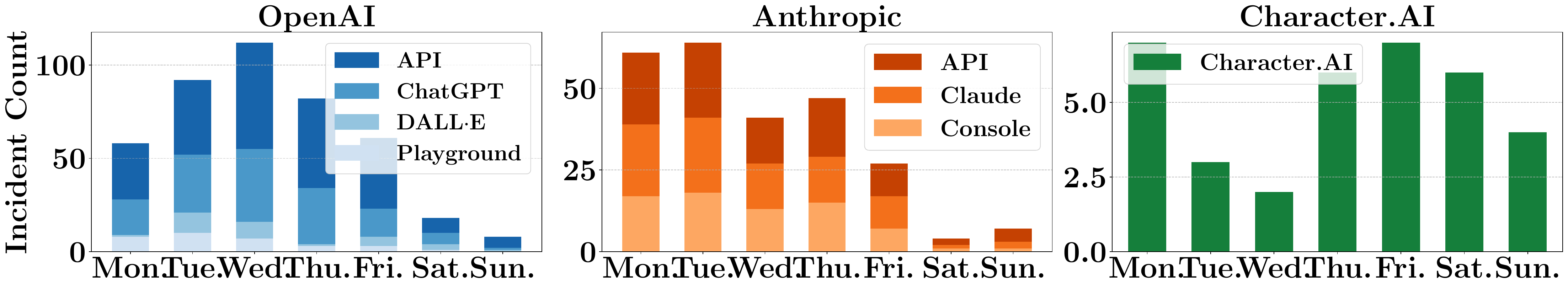}
        \vspace{-0.3cm}
        \caption{The number of incidents by day of week.}
        \label{fig:incident-day-of-week}
    \end{subfigure}
    \end{minipage}
    \begin{minipage}{\linewidth}
    \begin{subfigure}[b]{\linewidth}   
        \centering 
        \includegraphics[width=0.82\linewidth]{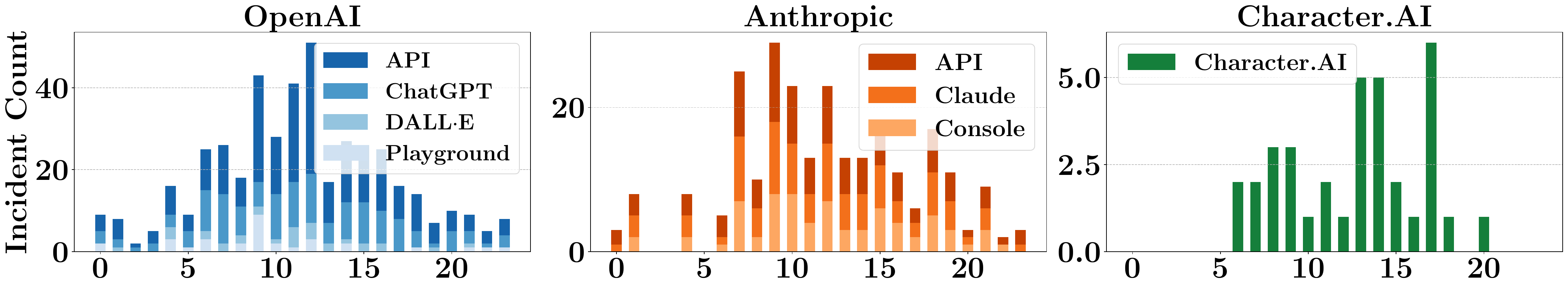}
        \vspace{-0.3cm}
        \caption{The number of incidents by hour of day.}
        \label{fig:incident-hour-of-day}
    \end{subfigure}
    \end{minipage}
  \vspace{-0.3cm}  
  \caption{Temporal distributions for incidents, PDT time.}
  \vspace{-0.3cm} 
  \label{fig:temporal-distribution}
\end{figure*}


To understand how the MTTR and MTBF values are distributed and compare the distributions for different LLM providers, \Cref{fig:mttr-and-mtbf-provider} displays the Empirical Cumulative Distribution Function (ECDF) plot of $MTTR$ in hours and $MTBF$ in days, grouped by provider. It also marks vertically different time points for better observation and comparison. 

A small percentage of incidents can be resolved within 10 minutes, such as 8.55\% for Anthropic.
Anthropic also solved the highest percent of incidents (37.18\%) in 0.5 hours, significantly more than OpenAI (19.72\%) and Character.AI (22.86\%). 
Most failures are addressed within 3 hours, with 74.25\% for OpenAI, 82.91\% for Anthropic, and 68.57\% for Character.AI. 
After 10 hours, 92.34\% of OpenAI, 90.60\% of Anthropic, and 91.43\% of Character.AI's failures are solved. 
However, a small proposition of failures for all providers lasted over 1 day, with 6.03\%, 7.69\%, and 5.71\%, respectively.
Overall, Anthropic resolved failures more quickly, despite a higher percentage of extreme cases lasting over 1 day.

Although Anthropic resolves failures the fastest, it also encounters them most frequently, with every 5.22 days on average. In contrast, OpenAI and Character.AI are more reliable, with failure occurring every 8.48 and 8.74 days, respectively.
A notable percentage of incidents occur within a day: 35.47\% for Anthropic, 28.77\% for OpenAI, and 20.00\% for Character.AI.
Within 1 week interval, nearly three-quarters of failures occur for OpenAI (75.64\%) and for Anthorpic (78.63\%), with a slightly lower rate for Character.AI (60.00\%). 
Over 90\% of incidents for all providers happen within a month of each other.
Our findings indicate that users of LLMs should expect failures regularly (at least once a month). Therefore, failure should not be an exceptional event but should be baked into the users' normal operating procedure.

\section{Failure Patterns Over Time}\label{sec:outage-analysis}\label{sec:uptime-analysis}\label{sec:analysis:temporal}
This section conducts time series analyses to examine the failure patterns over time, including: (1) Weekly and daily incident distributions, (2) Auto-correlations in different time intervals; and (3) Daily service available time.

\begin{figure}[t]
  \centering
  \begin{minipage}{\textwidth}
  \begin{subfigure}[b]{0.5\textwidth}
    \includegraphics[width=0.9\linewidth]{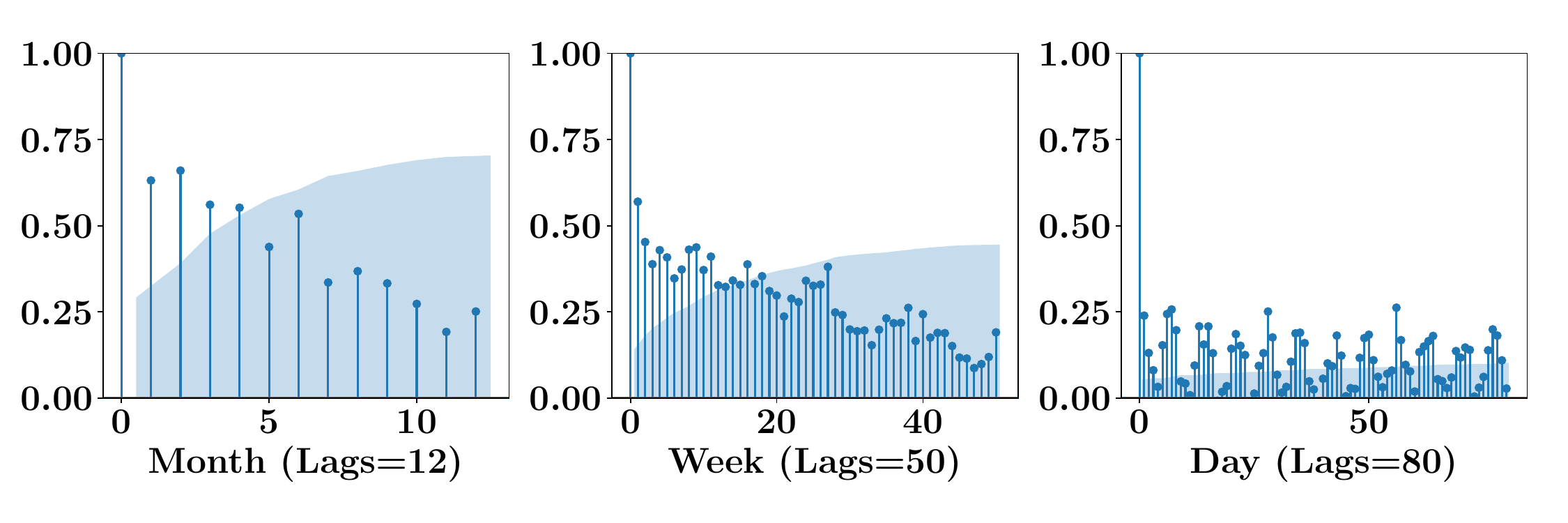}
    \vspace*{-0.3cm}
    \caption{OpenAI.}
    \label{fig:openai-acf}
  \end{subfigure}
  \vskip\baselineskip
  \vspace*{-0.3cm}
  \begin{subfigure}[b]{0.5\textwidth}
    \includegraphics[width=0.9\linewidth]{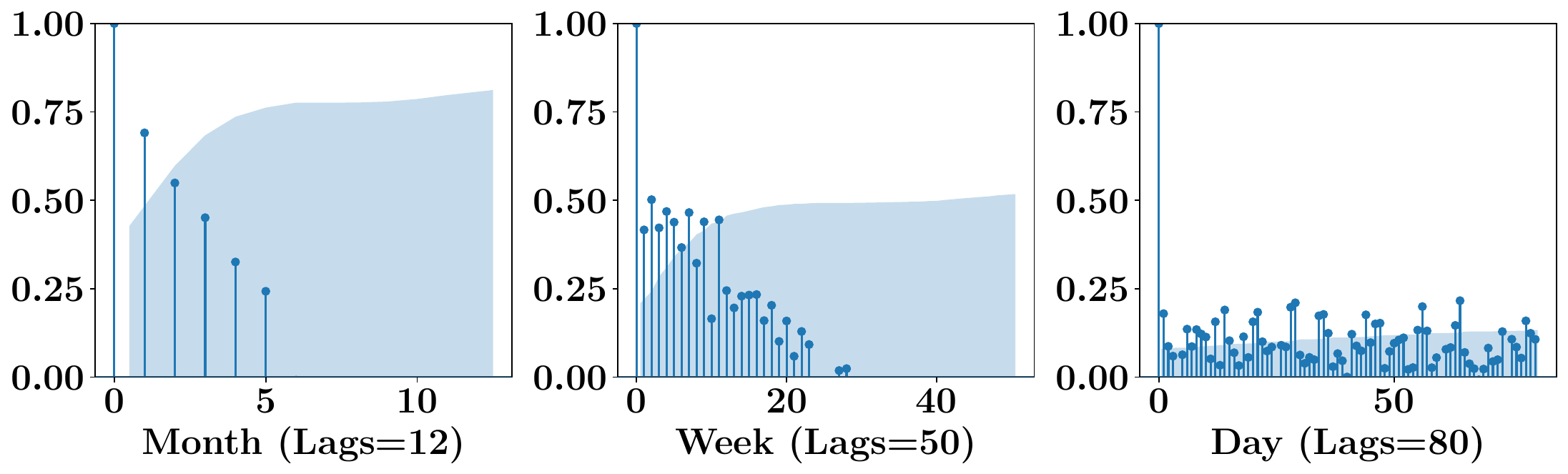}
    \vspace*{-0.3cm}
    \caption{Anthropic.}
    \label{fig:anthropic-acf}
  \end{subfigure}
  \end{minipage}
  \vspace{-0.3cm}
  \caption{Auto-correlations with the numbers of incidents aggregated at different time granularities.}
  \vspace{-0.5cm}
  \label{fig:auto-correlation}
\end{figure}

\subsection{Temporal Distributions}
\label{sec:analysis:temporal_dist}


\noindent\fbox{%
\parbox{\linewidth}{%
\defobservation{ob:time-dist}{OpenAI and Anthropic exhibit more failures on weekdays, while Character.AI has fewer on Tuesdays and Wednesdays. All services show peak failures from 8:00 to 16:00.}
}}

To investigate the temporal distributions of LLM incidents, we aggregate service incidents by day of week in \Cref{fig:incident-day-of-week}, and hour of day in \Cref{fig:incident-hour-of-day}. Incident times are given in local time (PDT) as they were originally reported in PDT.
OpenAI and Anthropic's services display a clear weekday pattern in incidents, with significantly more failures on weekdays than on weekends. In contrast, Character.AI follows a different pattern, with fewer failures occurring on Tuesdays and Wednesdays. 
This may be due to the differing purposes of using LLM services: Character.AI is primarily used for leisure \cite{character-ai-obsession}, while API and conversational services are more often used for work-related tasks, such as writing and coding \cite{DBLP:journals/corr/abs-2401-08329}.
All services exhibit a diurnal pattern, with incident peaks occurring during typical work hours, such as 8:00 to 16:00, and lower at night hours.
Similar periodic failure patterns are also found in machine learning jobs \cite{2023-hotcloudperf-mlfailures, chu2024genericmlworkloadshpc, DBLP:journals/fgcs/VersluisCGLPCUI23}, deep learning jobs \cite{DBLP:conf/hpca/LiASPABBRBHHHJK22}, and general user request in BurstGPT workloads \cite{wang2024burstgpt}.

\begin{figure*}[t]
  \centering
  \includegraphics[width=\linewidth]{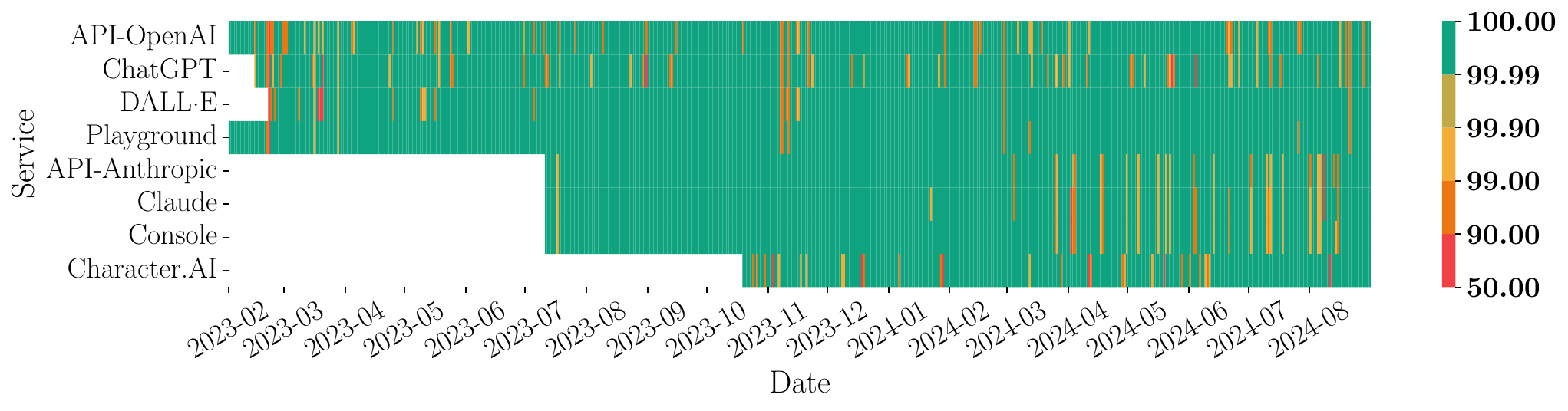}
  \vspace{-0.7cm}
  \caption{Service daily availability by scaled outage minutes [\%]. Some services started reporting later (see \Cref{sec:dataset}).}
  \label{fig:daily-availability}
    \vspace{-0.3cm}
\end{figure*}

\begin{table*}[t]
\centering
\caption{Service availability by scaled outage minutes (from all periods).}
\vspace{-0.3cm}
\label{tab:service-availability}
\resizebox{0.839\linewidth}{!}{
\begin{tabular}{rrrrrrrrrrrr}
\toprule
\textbf{Service} & \textbf{Min} & \textbf{Max} & \textbf{Mean} & \textbf{Median} & \textbf{=100\%} & \textbf{>=99.999\%} & \textbf{>=99.99\%} & \textbf{>=99.9\%} & \textbf{>=99\%} & \textbf{>=90\%} & \textbf{<90\%} \\
\midrule
API-OpenAI & 80.57\% & 100.0\% & 99.82\% & 100.0\% & 92.68\% & 92.68\% & 92.68\% & 92.84\% & 95.30\% & 99.77\% & 0.23\% \\
ChatGPT & 77.15\% & 100.0\% & 99.66\% & 100.0\% & 88.85\% & 88.85\% & 88.85\% & 89.20\% & 93.10\% & 99.12\% & 0.88\% \\
DALL·E & 74.17\% & 100.0\% & 99.78\% & 100.0\% & 95.88\% & 95.88\% & 95.88\% & 95.88\% & 96.77\% & 99.28\% & 0.72\% \\
Playground & 82.57\% & 100.0\% & 99.94\% & 100.0\% & 98.08\% & 98.08\% & 98.08\% & 98.16\% & 98.88\% & 99.84\% & 0.16\% \\
\midrule
API-Anthropic & 85.44\% & 100.0\% & 99.92\% & 100.0\% & 94.02\% & 94.02\% & 94.02\% & 94.26\% & 97.61\% & 99.76\% & 0.24\% \\
Claude & 82.02\% & 100.0\% & 99.84\% & 100.0\% & 93.06\% & 93.06\% & 93.06\% & 93.06\% & 97.13\% & 99.52\% & 0.48\% \\
Console & 82.56\% & 100.0\% & 99.89\% & 100.0\% & 93.54\% & 93.54\% & 93.54\% & 93.54\% & 97.61\% & 99.76\% & 0.24\% \\
\midrule
Character.AI & 85.33\% & 100.0\% & 99.59\% & 100.0\% & 90.88\% & 90.88\% & 90.88\% & 91.19\% & 94.03\% & 98.11\% & 1.89\% \\
\bottomrule
\end{tabular}}
    \vspace{-0.3cm}
\end{table*}

\subsection{Auto-correlations}

\vspace*{0.06cm}
\noindent\fbox{%
\parbox{\linewidth}{%
\defobservation{ob:auto-correlation}{LLM service failures have strong monthly auto-correlations, with OpenAI incidents showing longer-lasting correlations than Anthropic. Both services display distinct weekly periodicity.}
}}
\vspace*{0.03cm}

We investigate if a failure is immediately followed by another failure and how often it happens. \Cref{fig:auto-correlation} depicts the auto-correlation for the number of incidents at month, week, and day granularities.
We use the autocorrelation function (ACF) to measure the degree of correlation based on the temporal incidents data. Confidence intervals are drawn as the blue area. By default, this is set to a 95\% confidence interval, suggesting that correlation values outside of this area are significant, which are real patterns rather than random noise. Lags represent the time intervals at which a time series is compared to itself, and autocorrelation measures how similar a time series is to itself at different lags.

For OpenAI, the auto-correlation plots display significant positive correlations up to lag 3 on a monthly scale and up to lag 12 on a weekly scale, indicating that both monthly and weekly incidents are strongly related to their previous values. Anthropic shows similar correlations with shorter lags, with up to lag 1 for monthly data and lag 7 for weekly data, likely affected by Anthropic's shorter operational history.
The consistent but gradual decay in auto-correlations at every 7-day interval for both OpenAI and Anthropic suggests strong weekly periodic behavior, supporting our previous findings in \Cref{fig:temporal-distribution}. Compared to the auto-correlations observed in ML failures from the previous study \cite{2023-hotcloudperf-mlfailures}, the auto-correlation in LLM service failures shows stronger periodic trends. The periodic characteristics can be utilized to predict future incidents, similar to workload failure predictions \cite{DBLP:conf/IEEEcloud/LiWADSC23}.

\subsection{Service Availability Over Time}

\vspace*{0.06cm}
\noindent\fbox{%
\parbox{\linewidth}{%
\defobservation{ob:availability}{
ChatGPT is the least consistently available service, with only 88.85\% of days fully available, followed by Character.AI at 90.88\%. 
Availability of Anthropic's services declined after April 2024, possibly due to product release and the sharp increase in user demands.}
}}
\vspace*{0.03cm}



We provide a high-level view of what level of service reliability a user can expect in this section.
\Cref{fig:daily-availability} shows the service daily availability by scaled outage minutes, from February 2023 to August 2024. 
We categorized availability into five levels based on their value ranges. Days without outages, which mean full service availability, are colored green, while days with longer outage durations are represented by colors closer to red.
\Cref{tab:service-availability} gives the specific statistics of service availability.
DALL·E and Playground have the highest availability, with 95.88\% and 98.08\% of days fully accessible, respectively. In contrast, ChatGPT is the least available service, with only 88.85\% of days fully accessible. 
Availability of Anthropic's services declined after April 2024, possibly due to product release and the sharp increase in user demands \cite{claude-release-notes, api-release-notes}. 
Character.AI also shows noticeable instability, with only 90.88\% of days fully available and over 1.89\% of days with availability falling below 90\%.


\section{Co-occurrence of Failures}\label{sec:concurrent-analysis}\label{sec:concurrent}\label{sec:analysis:coocurrence}
This section examines the co-occurrence of failures across services. When an outage occurs in one service, do other services also experience outages? Is there any co-occurrence within and across different providers, given that services may share the same cloud infrastructure? For instance, both Anthropic and Character.AI rely on GCP \cite{gcp-claude, gcp-characterai}. How about the impacted range of incidents for different services and providers? To address these questions, we analyze (1) the co-occurrence of outages, and (2) the impact range of incidents.

\subsection{Co-occurrence of Outages}

\begin{figure*}[t]
  \centering
  \begin{minipage}{\textwidth}
  \begin{subfigure}[b]{0.5\textwidth}
    \includegraphics[width=\linewidth]{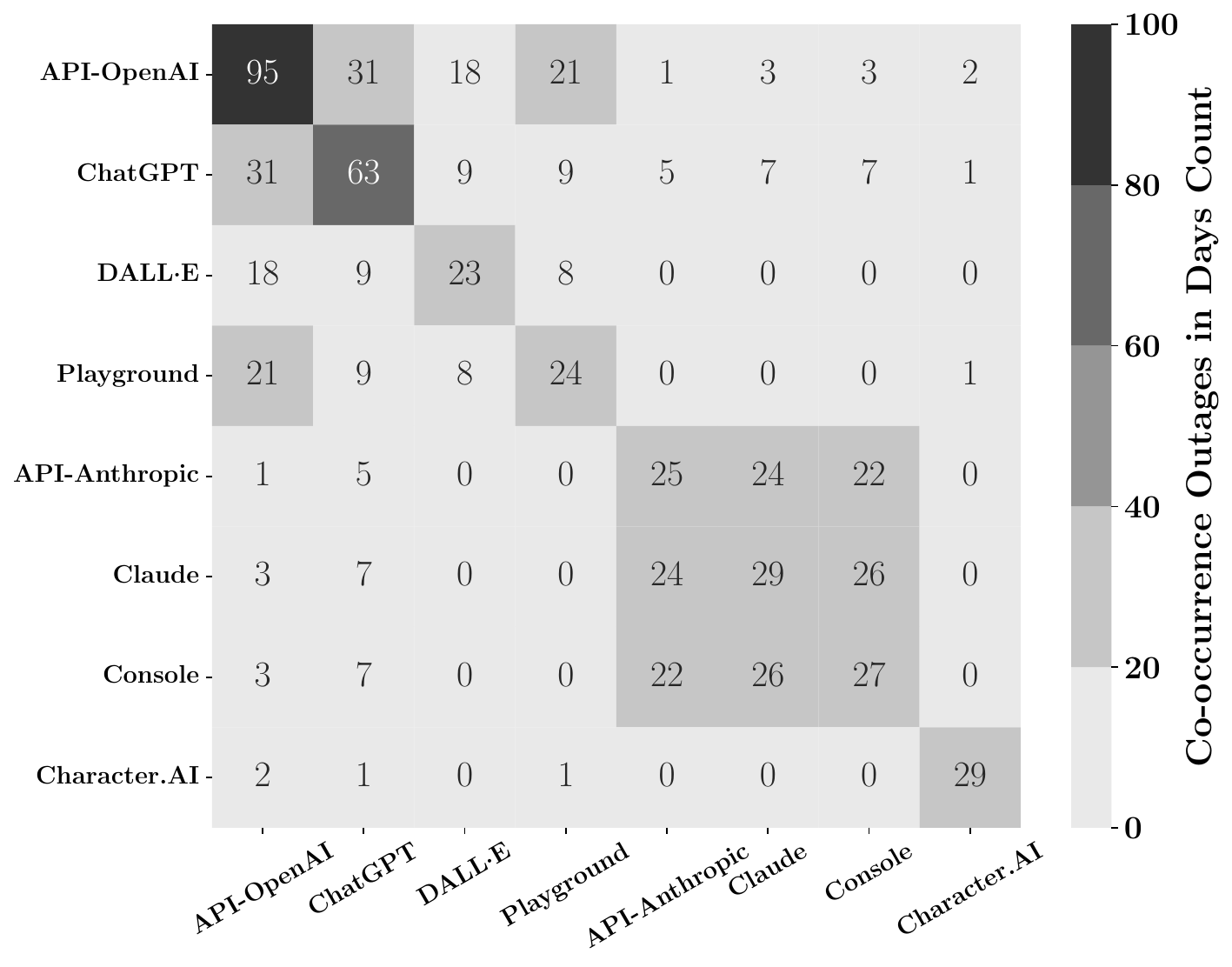}
        \vspace*{-0.3cm}
    \caption{Co-occurrence outages in days count.}
    \label{fig:outage-count-heatmap}
  \end{subfigure}
  \begin{subfigure}[b]{0.5\textwidth}
    \includegraphics[width=\linewidth]{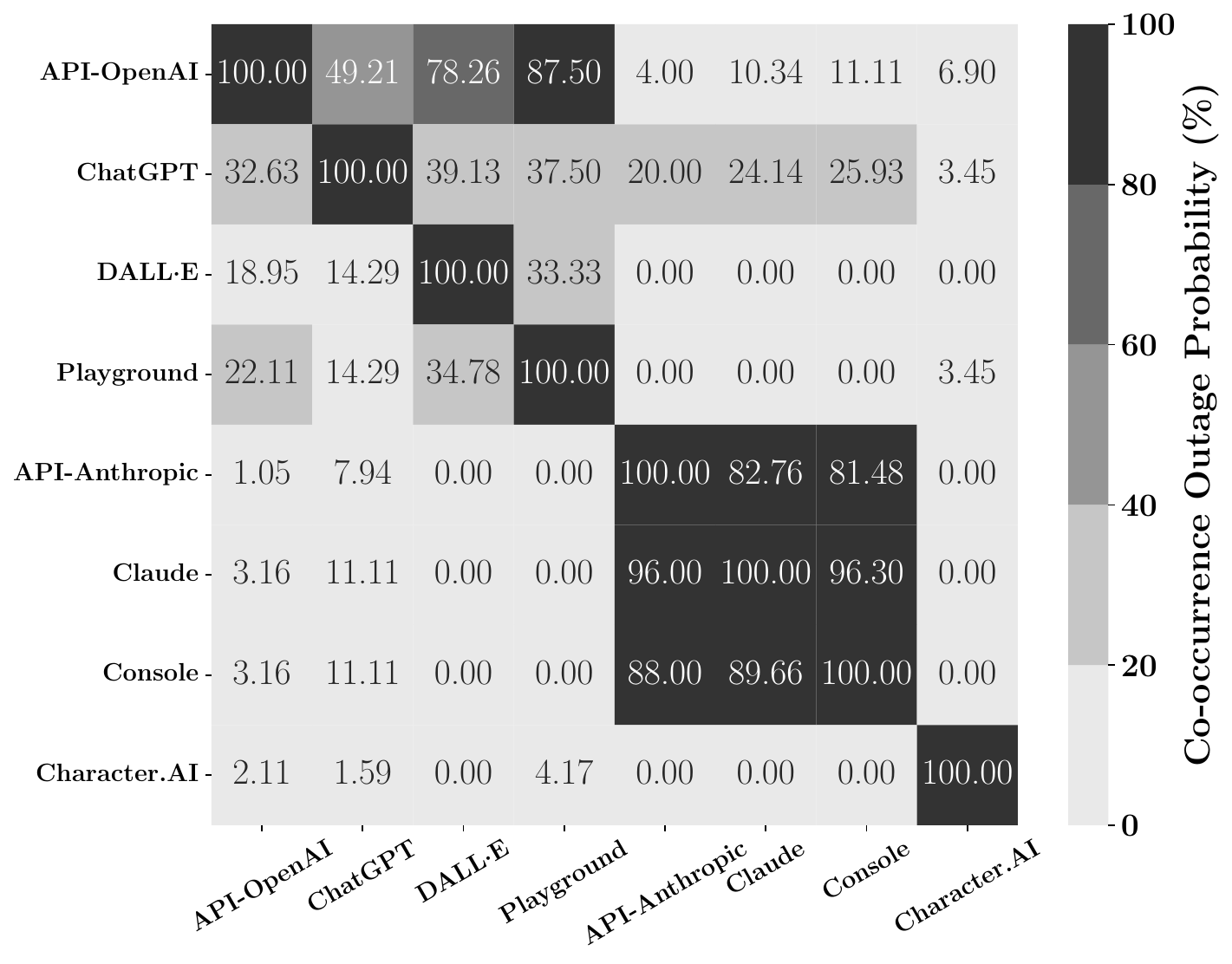}
        \vspace*{-0.7cm}
    \caption{Conditional probabilities of co-occurrence outages. Notes: y-axis $=$ service A, x-axis $=$ service B, cells $=$ P(A|B).}
    \label{fig:outage-prob-heatmap}
  \end{subfigure}
  \end{minipage}
  \vspace{-0.3cm}
  \caption{Co-occurrence of outages between service pairs.}
  \vspace{-0.3cm}
  \label{fig:concurrent-outages}
\end{figure*}

\vspace*{0.06cm}
\noindent\fbox{%
\parbox{\linewidth}{%
\defobservation{cooccur-outage}{Co-occurrence is particularly high among services from the same provider, suggesting a strong interdependence between those services.
For Anthropic’s services, the likelihood of any two services experiencing outages on the same day is over 80\%, indicating a severe lack of isolation across different services.}
}}
\vspace*{0.03cm}


The \Cref{fig:outage-count-heatmap} shows the number of co-occurring outages across different services on the same day. 
The counts of outages may be affected by the maximum number of outages. For example, the number of co-occurrences among Anthropic services is lower than for OpenAI services, however, the probability of co-occurrence among Anthropic services is higher.
To avoid this impact of the number of outages, we also give the conditional probabilities of co-occurring outages in \Cref{fig:outage-prob-heatmap}. 
The conditional probability indicates the likelihood that if service B experiences an outage, service A will also experience an outage. For instance, the 49.21\% in row 1, column 2 means that if ChatGPT is down, there is a 49.21\% chance that OpenAI's API will also experience an outage on the same day. The probability that service A is also outage while service B is outage can be formulated in:
\begin{equation}
    P(A|B) = \frac{P(A \cap B)}{P(A) \times P(B)} = \frac{O_{AB}}{O_B}
\end{equation}
$O_{AB}$ represents the number of days when both services A and B experience outages simultaneously, while $O_B$ indicates the number of days that service B has an outage.

The heatmaps show that co-occurrence is notably high among services from the same provider. For OpenAI services, the API is more likely to have an outage with DALL·E (78.26\%) and Playground (87.50\%) than ChatGPT (49.21\%).
For Anthropic’s services, the likelihood of any two services experiencing outages on the same day is extremely high over 80\%, this may be caused by a lack of isolation across different services.
There is no correlation observed between services from different providers. The lack of correlation suggests that user can use one service as the other's backup to increase their reliability.
The difference in co-occurrence between OpenAI and Anthropic suggests that outages could be reduced through better service isolation. Based on publicly available information, this may be caused by the different cloud infrastructures used by LLMs: OpenAI relies on Azure \cite{azure-openai}, while Anthropic uses GCP \cite{gcp-claude}.

\subsection{Impact Range of Incidents}


\begin{figure}[t]
  \centering
  \begin{minipage}{\textwidth}
  \begin{subfigure}[b]{0.25\textwidth}
    \includegraphics[width=\linewidth]{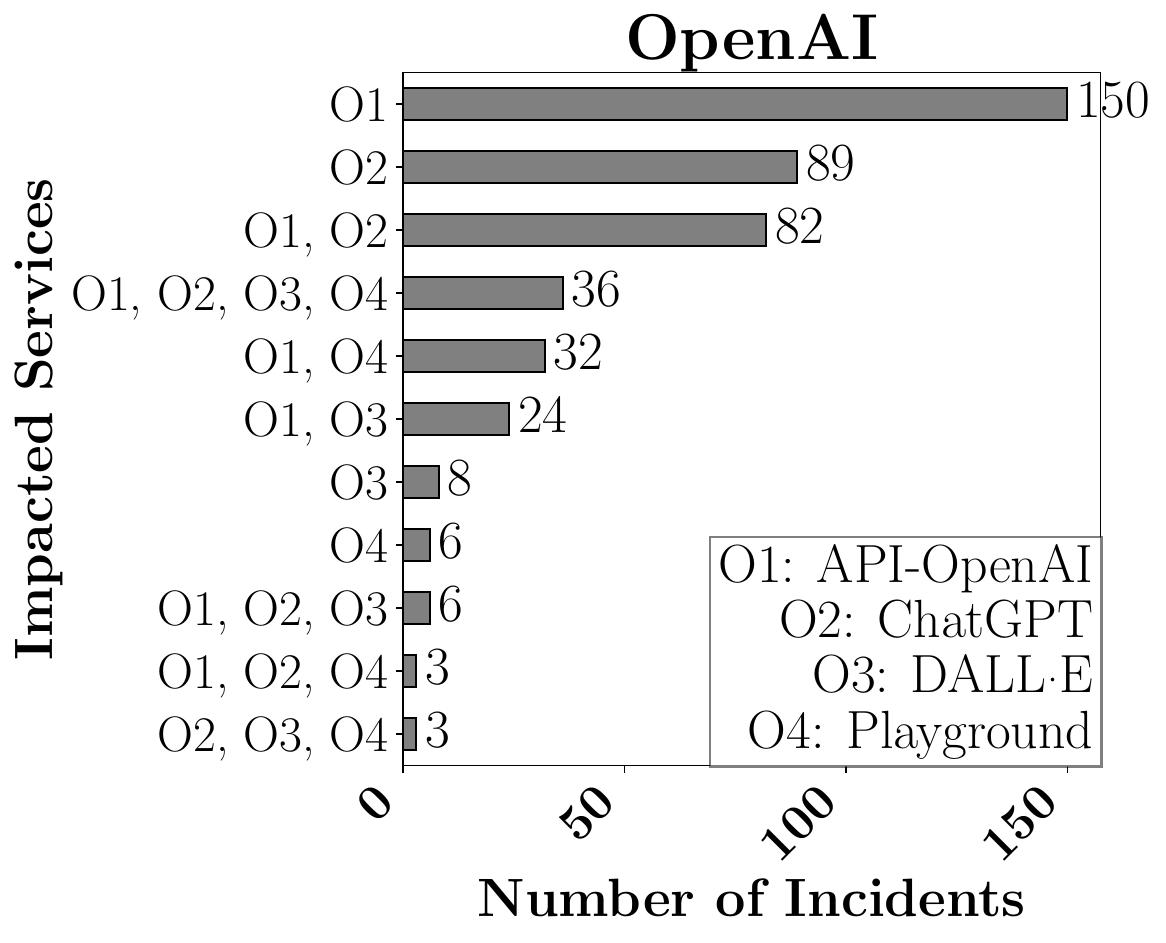}
        \vspace*{-0.5cm}
  \end{subfigure}
  \begin{subfigure}[b]{0.23\textwidth}
    \includegraphics[width=\linewidth]{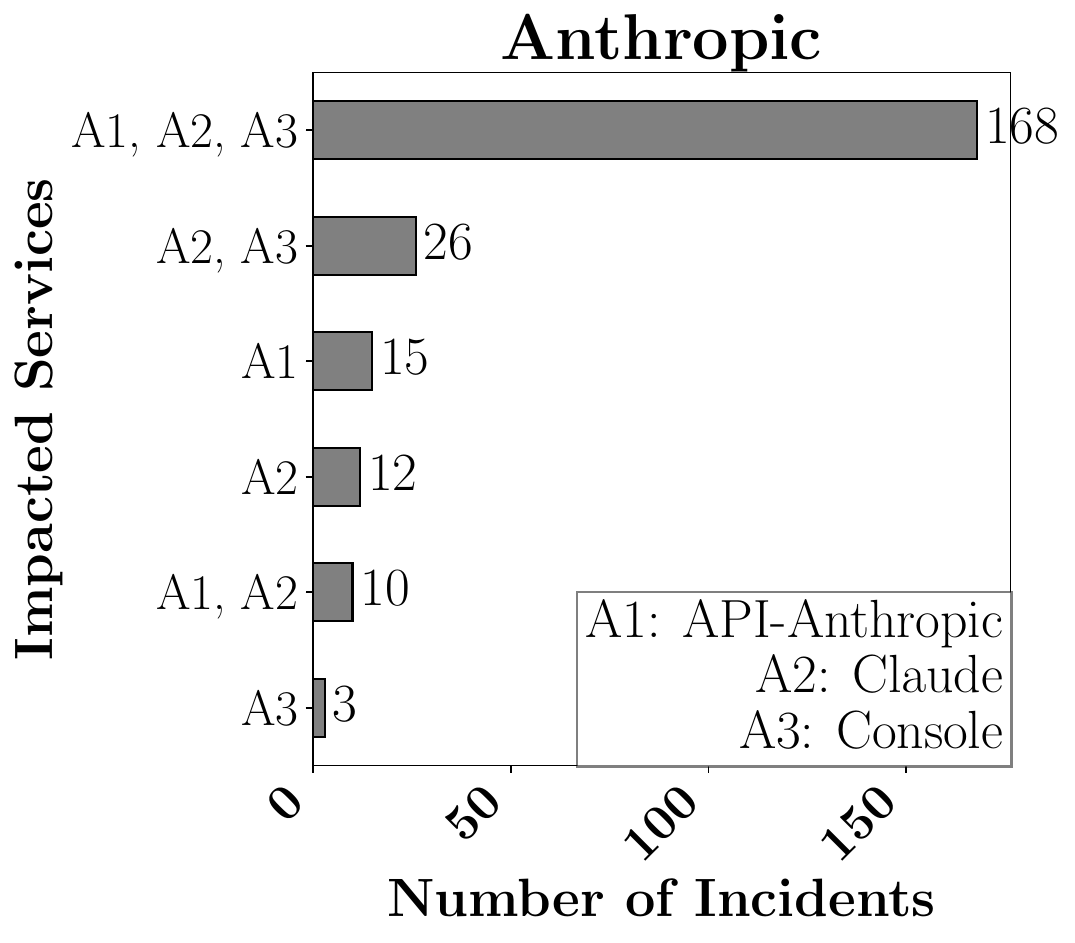}
        \vspace*{-0.5cm}
  \end{subfigure}
  \end{minipage}
  \vspace{-0.3cm}
  \caption{Impact Services of OpenAI and Anthropic incidents, respectively.}
  \vspace*{-0.3cm}
  \label{fig:impact-range}
\end{figure}

\begin{table}[t]
\caption{Percentage of the number of impacted services.}
\vspace{-0.3cm}
\label{tab:impact-range-percent}
\resizebox{0.68\linewidth}{!}{
\begin{tabular}{lrrrr}
\toprule
\textbf{ } & \textbf{1} & \textbf{2} & \textbf{3} & \textbf{4} \\
\midrule
OpenAI & 57.63\% & 31.44\% & 2.73\% & 8.20\% \\
Anthropic & 12.82\% & 15.38\% & 71.79\% & - \\
\bottomrule
\end{tabular}}
\vspace{-0.3cm}
\end{table}

\vspace*{0.06cm}
\noindent\fbox{%
\parbox{\linewidth}{%
\defobservation{cooccur-incident}{71.79\% of Anthropic incidents affect all its services, compared to only 8.20\% for OpenAI.}
}}
\vspace*{0.03cm}

The incident reports indicate that one incident can impact several services, which is the impact range of incidents. \Cref{fig:impact-range} gives the impacted services combinations of OpenAI and Anthropic incidents based on their reports, respectively.
For Anthropic, the majority of incidents (71.79\%) impact its 3 services jointly. However, for OpenAI, only 8.20\% of services are impacted together, and over half (57.63\%) of incidents affect a single service.

\section{Limitations and Validity}\label{sec:threats}\label{sec:scope}\label{sec:validity}

The \textit{generality} of our work is limited to the services we analyze. We analyze LLM-related services from three popular operators, including the currently most popular (OpenAI). However, other, very different services could exist, e.g., allowing users to self-host LLMs (e.g., Anyscale), and LLM services from big cloud providers and/or used primarily internally (e.g., Google Gemini).

The \textit{accuracy} of our failure dataset is limited to what the LLM operators themselves report. That makes our data subject to the operators' bias. Prior work~\cite{DBLP:conf/nsdi/HuZYMW20, DBLP:conf/acsos/TalluriOVTI21} suggests the operator's reports already capture the most user-visible failures as those generate widespread social media coverage, making it difficult for the operators to hide failure; to confirm this for LLM services, we need to collect data from other sources such as the user devices~\cite{DBLP:conf/nsdi/BurnettCCEGJMPR20} or user failure reports~\cite{DBLP:conf/nsdi/HuZYMW20, DBLP:conf/acsos/TalluriOVTI21}.

The \textit{depth} of our analysis regarding the root cause of failures is limited. We glean limited information from the operators' failure reports regarding the hardware and software infrastructure. To confirm our findings, we would ideally use detailed infrastructure and application-level data. 
However, this requires active help from the service operators, e.g., releasing their system traces as Google did with its cluster workloads~\cite{google-cluster-data}.

The \textit{scope} of our work is limited to LLM services. We ignore other deep learning services such as Image Generation (e.g., Stable Diffusion, Midjourney), Translation (e.g., DeepL), etc. However, we believe the operational characteristics of LLM services are valuable in and of themselves. LLMs have gained broad general public adoption and mindshare, as described in \Cref{sec:intro}. LLM services now also support multi-modal use cases such as image generation and image-based question answering, making them some of the most general deep learning tools currently available.




\section{Related Work}\label{sec:related-work}

Overall, this work complements the existing body of work on failure characterization, modeling, and more generally failure-recovery, with a focus on the emerging area of LLM services. Ours is the first comprehensive, longitudinal data collection and empirical characterization of public LLM services.

\noindent\textbf{Operational failure characterization} of workstations~\cite{DBLP:journals/jpdc/JavadiKIE13}, HPC sites~\cite{DBLP:conf/sc/GuptaPET17}, clouds~\cite{DBLP:conf/hase/GarraghanTX14}, big data jobs~\cite{DBLP:conf/ccgrid/RosaCB15}, networks~\cite{DBLP:conf/cloud/PotharajuJ13}, storage devices~\cite{DBLP:conf/sc/AlterXDS19}, CPUs~\cite{DBLP:conf/hotos/HochschildTMGRC21}, and GPUs~\cite{DBLP:conf/hpca/TiwariGRMRVOLDN15} has led to improved application designs and fault-tolerance mechanisms. Leading from these, we have better failure detection~\cite{DBLP:conf/nsdi/BurnettCCEGJMPR20, DBLP:conf/nsdi/HuZYMW20}, checkpointing~\cite{DBLP:conf/dsn/GargPCT18}, retry~\cite{DBLP:conf/nsdi/PrimoracAB21}, and replication mechanisms~\cite{DBLP:conf/ccgrid/ShenIICRE15}. However, these do not cover 
deep learning and particularly LLM services. 

There is existing work on operational characteristics of GPUs for deep learning~\cite{DBLP:conf/sc/LiHSTPEK17}, ML jobs on HPC clusters~\cite{chu2024genericmlworkloadshpc}, and deep-learning clusters~\cite{DBLP:conf/usenix/JeonVPQXY19}. However, no work has described the 
\textit{operational failure characteristics} of user-facing deep learning services. Our study addresses this gap, focusing on LLMs. 

\noindent\textbf{Deep learning workloads} have been characterized including their GPU utilization~\cite{DBLP:conf/sc/Hu0Y0021, DBLP:conf/hpca/LiASPABBRBHHHJK22}, network characteristics~\cite{DBLP:journals/micro/AwanJCSP20}, and storage characteristics~\cite{DBLP:conf/sc/ChienMSSHNL18}. User-facing machine learning workloads have also been characterized~\cite{DBLP:conf/nsdi/WengXYWWHLZLD22}. The studies complement our work as they explore different hardware/software stack layers. We complement the studies by enhancing the community's understanding of LLM failures at the user-facing application layer.

\noindent\textbf{LLM workloads} have been characterized at the preliminary-level for training~\cite{DBLP:conf/nsdi/Hu0WWZC0L0L0024}, fine-tuning~\cite{DBLP:journals/corr/abs-2408-04693Xia, wang2024burstgpt}, and inference~\cite{lazuka2024llm}. 
Failures have been assessed briefly; e.g., found to occur frequently ($\sim$9~hour MTBF) in LLM training~\cite{DBLP:conf/nsdi/Hu0WWZC0L0L0024}, compared to around 4~days MTBF for the user-facing services in this work. 
Fine-tuning and inference workloads have not been characterized, especially concerning failures.
Ours is the first study to focus on failures occurring in public LLM services, with unique contributions in longitudinal analysis and in collecting comprehensive data from multiple services.


\section{Conclusion}\label{sec:conclusion}

Understanding the characteristics of failures in the operation of public LLM services has become a stringent problem, driven by the rapid increase in the popularity of such services, market competitiveness, and increasingly self-reported presence of such failures by LLM service providers. 
Addressing this problem, in this work we have conducted a comprehensive empirical characterization of long-term outages and incidents in public LLM services.

Our main findings includes: (1) Different LLM services take varying amounts of time at different stages of the failure-recovery process. For example, Anthropic services spent more time for investigating and resolving issues than OpenAI services. (2) OpenAI and Anthropic's services failures exhibit periodic patterns that are more frequent on weekdays than on weekends. However, Character.AI has fewer failures on Tuesdays and Wednesdays. (3) Co-occurrence is significantly higher among services within the same provider, with no clear co-occurrence observed between services from different providers.

Overall, we emphasized over 10 observations, which scientists, engineers, and users could include directly in their knowledge base, and from which improvements to LLM systems could occur in time.
For the future work, we aim to lead a community effort where LLM service availability datasets, collected long-term and processed to provide similar information, can be shared. Future analysis could include promising emerging LLM services in different countries such as Google Gemini, Mistral AI, and DeepSeek.



\begin{acks}
This work was supported by the EU Horizon Graph Massivizer and the EU MSCA Cloudstars projects. This research was partly supported by a National Growth Fund through the Dutch 6G flagship project "Future Network Services".
We thank the China Scholarship Council (CSC) for supporting Xiaoyu Chu. 
We thank Daniel Hofstätter, Krijn Doekemeijer, and Radu Nicolae for their critical review of the manuscript.
\end{acks}

\bibliographystyle{ACM-Reference-Format}
\bibliography{reference}



\end{document}